\documentclass[aps,a4paper,floatfix,nofootinbib]{revtex4}

\usepackage{graphicx,color}
\usepackage{amsmath,amssymb,amsfonts,amsthm,amscd,bm}
\usepackage{booktabs}
\usepackage{cancel}
\usepackage{accents}

\def\tilde{\widetilde}
\def\bar{\overline}
\def\hat{\widehat}
\def\*{\star}
\def\[{\left[}
\def\]{\right]}
\def\({\left(}      

\def\){\right)}

\def\frac#1#2{\dfrac{#1}{#2}}
\def\inv#1{\dfrac{1}{#1}}
\def\half{\tfrac{1}{2}}
\def\d{\partial}

\def\ket#1{ | #1 \rangle}

\def\2pi{\hbox{$2\pi i$}}

\def\dsl{\raise.15ex\hbox{/}\kern-.57em\partial}
\def\Dsl{\,\raise.15ex\hbox{/}\mkern-.13.5mu D}
      \def\CC{{\cal C}}
      
   \def\CH{{\cal H}}   
   \def\CK{{\cal K}}   
\def\CM{{\cal M}}   \def\CN{{\cal N}}   \def\CO{{\cal O}}
\def\CP{{\cal P}}   \def\CQ{{\cal Q}}   
\def\CS{{\cal S}}   \def\CT{{\cal T}}   \def\CU{{\cal U}}

\def\2pi{\hbox{$2\pi i$}}

\def\dsl{\raise.15ex\hbox{/}\kern-.57em\partial}
\def\Dsl{\,\raise.15ex\hbox{/}\mkern-.13.5mu D}
\font\numbers=cmss12
\font\upright=cmu10 scaled\magstep1
\def\stroke{\vrule height8pt width0.4pt depth-0.1pt}
\def\topfleck{\vrule height8pt width0.5pt depth-5.9pt}
\def\botfleck{\vrule height2pt width0.5pt depth0.1pt}
\def\Zmath{\vcenter{\hbox{\numbers\rlap{\rlap{Z}\kern
    0.8pt\topfleck}\kern 2.2pt
    \rlap Z\kern 6pt\botfleck\kern 1pt}}}
\def\Qmath{
    \vcenter{\hbox{\upright\rlap{\rlap{Q}\kern3.8pt\stroke}\phantom{Q}}}}
\def\Nmath{\vcenter{\hbox{\upright\rlap{I}\kern 1.7pt N}}}
\def\Cmath{\vcenter{\hbox{\upright\rlap{\rlap{C}\kern
                   3.8pt\stroke}\phantom{C}}}}
\def\Rmath{\vcenter{\hbox{\upright\rlap{I}\kern 1.7pt R}}}

\def\Xmath{\vcenter{\hbox{\numbers\rlap{\rlap{X}\kern
    0.8pt\topfleck}\kern 2.2pt
    \rlap X\kern 6pt\botfleck\kern 1pt}}}

\def\Z{\ifmmode\Zmath\else$\Zmath$\fi}
\def\Q{\ifmmode\Qmath\else$\Qmath$\fi}
\def\N{\ifmmode\Nmath\else$\Nmath$\fi}
\def\C{\ifmmode\Cmath\else$\Cmath$\fi}
\def\R{\ifmmode\Rmath\else$\Rmath$\fi}
\def\barray{\begin{eqnarray}}
\def\earray{\end{eqnarray}}
\def\beq{\begin{equation}}
\def\eeq{\end{equation}}

\def\kvec{{\bf{k}}}
\def\xvec{{\bf{x}}}

\def\AA{\leavevmode\setbox0=\hbox{h}
\dimen0=\ht0 \advance\dimen0 by-1ex\rlap{\raise.67\dimen0\hbox{\char'27}}A}

\makeatletter
\def\iddots{\mathinner{\mkern1mu\raise\p@
\vbox{\kern7\p@\hbox{.}}\mkern2mu
\raise4\p@\hbox{.}\mkern2mu\raise7\p@\hbox{.}\mkern1mu}}
\makeatother

\theoremstyle{plain}

\theoremstyle{remark}

\def\SU{{\rm SU} (N+1)}

\def\Z{\mathbb{Z}}

\def\K{\CK}

\def\daggerc{{\dagger_\kappa}}

\def\Phitilde{\tilde{\Phi}}

\def\Jtilde{\tilde{J}}

\def\K{\CK}

\def\ketket{\rangle\!\rangle}
\def\brabra{\langle\!\langle}

\def\paulix{\begin{pmatrix} 0 & 1 \\ 1 & 0 \end{pmatrix} }
\def\pauliy{\begin{pmatrix} 0 & -i  \\ i & 0 \end{pmatrix} }
\def\pauliz{\begin{pmatrix} 1 & 0 \\ 0 & -1 \end{pmatrix} }

\def\updoublet{\begin{pmatrix} u\\d \end{pmatrix}}
\def\charmdoublet{\begin{pmatrix} c\\s \end{pmatrix}}
\def\topdoublet{\begin{pmatrix} t\\b \end{pmatrix}}
\def\edoublet{\begin{pmatrix} \nu_e\\  e  \end{pmatrix}}
\def\muondoublet{\begin{pmatrix} \nu_\mu \\ \mu \end{pmatrix}}
\def\taudoublet{\begin{pmatrix} \nu_\tau \\ \tau \end{pmatrix}}

\def\X{{\bf{X}}}

\def\SU2{{\rm SU(2)}}

\def\SL2Z{{\rm SL}(2,\Z)}

\def\psiplus{\psi_{\plusdot}}
\def\psiminus{\psi_{\minusdot}}
\def\psipm{\psi_{\pmdot}}

\def\Pplus{P_{\plusdot}}
\def\Pminus{P_{\minusdot}}
\def\Ppm{P_{\pmdot}}

\def\Hplus{\CH_{\plusdot}}
\def\Hminus{\CH_{\minusdot}}
\def\Hpm{\CH_{\pmdot}}

\newcommand{\plusdot}{\mathbin{+\mkern-5mu\raise0.42ex\hbox{$\cdot$}}}
\newcommand{\minusdot}{\mathbin{-\mkern-5mu\raise0.42ex\hbox{$\cdot$}}}

\newcommand{\pmdot}{\mathbin{\pm\mkern-5mu\raise0.42ex\hbox{$\cdot$}}}
\newcommand{\mpdot}{\mathbin{\mp\mkern-5mu\raise0.42ex\hbox{$\cdot$}}}

\def\smallpsi{{\scriptscriptstyle \psi}}

\makeatletter
\newcommand{\dotminus}{\mathbin{\mathpalette\dotminus@aux{}}}
\newcommand{\dotminus@aux}[2]{%
  \ooalign{%
    $\m@th#1-$\cr
    \hidewidth$\m@th#1\cdot$\hidewidth
  }%
}
\makeatother

\def\qop{\CQ}

\def\Nanti{\bar{N}}

\def\Nbar{\Nanti}

\begin{document}

\title{A rich structure of renormalization group flows\\
for Higgs-like  models in 4 dimensions.}
\author{
 Andr\'e  LeClair\footnote{andre.leclair@cornell.edu} 
}
\affiliation{Cornell University, Physics Department, Ithaca, NY 14850,  USA} 

\begin{abstract}

We consider $2$  coupled Higgs doublets which transform in the usual way under SU(2).  
By constructing  marginal operators which satisfy an operator product expansion  based on the SU(2) Lie algebra,  we can obtain a rich pattern of 
renormalization group (RG)  flows which includes lines of fixed points and more interestingly,   cyclic RG flows which are 
unavoidable in this model.    
The hamiltonian is pseudo-hermitian,  $H^\dagger = \K H \K^\dagger $ with $\K$ unitary satisfying $\K^2 =1$,   thus the model is  
non-unitary.    The hamiltonian still has real eigenvalues,   but the non-unitarity is manifested in negative norm states. 
Based on a generalized optical theorem for pseudo-hermitian hamiltonians,  we show that our model is in fact unitary below the threshold for 
particle/anti-particle pair production.    It is thus unitary in the non-relativistic limit,  which opens up some potential applications to condensed matter physics.  
We argue  that our  model breaks $\CC\CP$ symmetry.   
Upon spontaneous symmetry breaking,   the Higgs-like  fields have an infinite number of vacuum expectation values $v_n$ 
which satisfy ``Russian Doll" scaling $v_n \sim e^{2 n \lambda}$ where $n=1,2,3,\ldots$ and $\lambda$ is the period of one
RG cycle which is an RG invariant.    We speculate that this Russian Doll RG flow can perhaps resolve the so-called hierarchy problem and may  shed light on  the origin of ``families" in the 
Standard Model of particle physics.     If after spontaneous symmetry breaking of the SU(2) to U(1) a cyclic RG with period $\lambda$  is operative up to the electro-weak scale,   then this admits 3 RG cycles,   i.e. 3 families of quarks and leptons.    The strongest constraints on the RG period $\lambda$ comes from the phenomenological Koide formula,   wherein 
$\lambda \approx \pi/2$.\footnote{{\bf Note added:}    While this article was under review,   we extended the 1-loop calculations in this article to higher orders,   which confirmed that  the main features of the RG flows considered in this paper persist \cite{RDolls2},   where in the latter article we left  aside the more speculative ideas on the hierarchy problem and the origin of families. 
In the present article  the focus is on the cyclic RG flows to 1-loop.        Below   we incorporate some of the results in \cite{RDolls2}   as {\bf ``Note added"}  footnotes.} 

\end{abstract}

\maketitle
\tableofcontents

\section{Introduction}

\begin{figure}[t]
\centering\includegraphics[width=.4\textwidth]{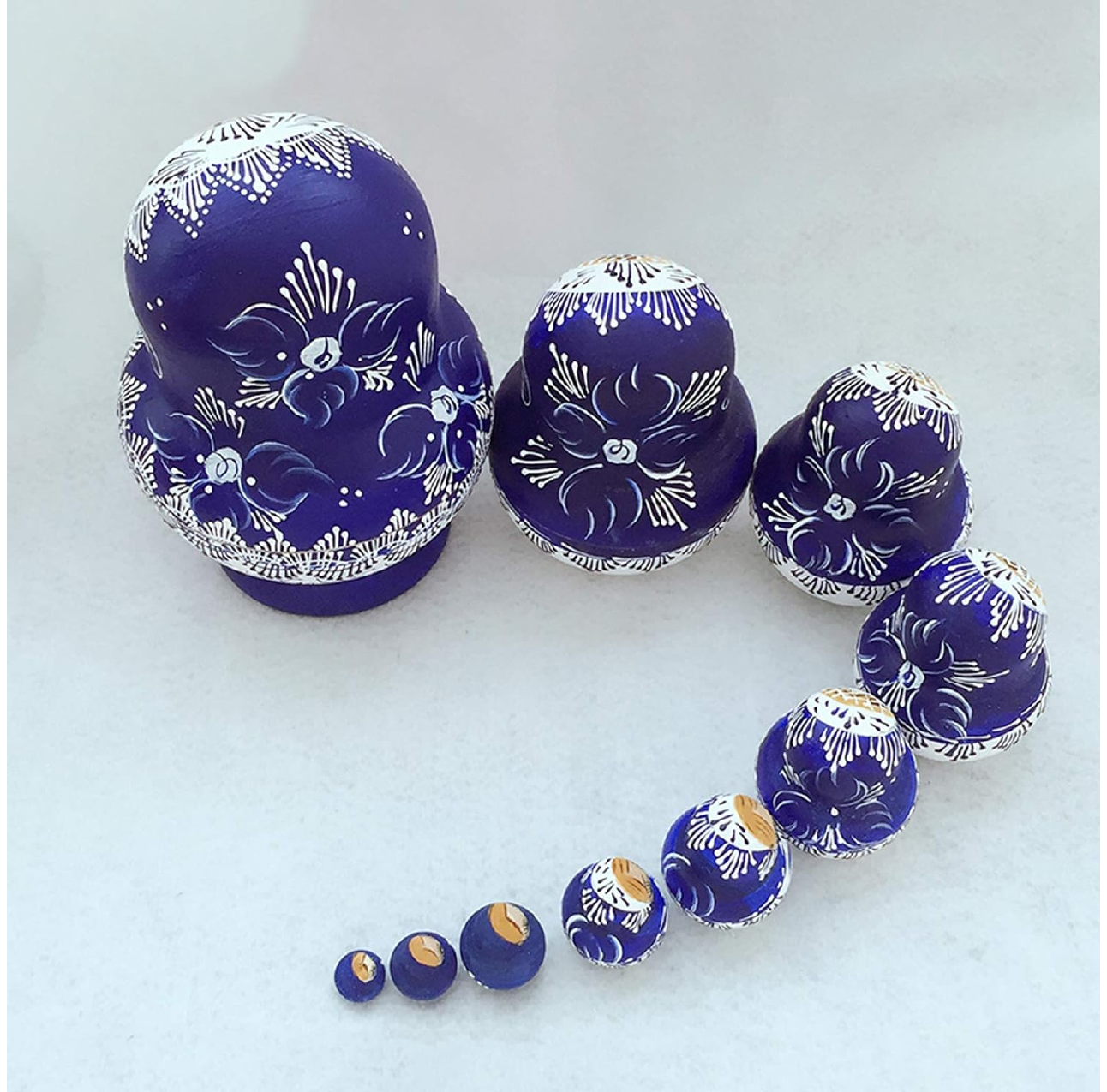}
\caption{Russian Doll craftmanship representing  10 families for 10 RG cycles.}
 \label{Dolls}
\end{figure}

In K.  Wilson's pioneering work on the renormalization group (RG),     he attempted to classify  possible RG flow behavior in a model independent way \cite{KWilson}.       Although asymptotic freedom,   which is  essential to understanding QCD \cite{Wilczek,Gross},    was curiously overlooked,   he did point out the possibility of more exotic  limit-cycle behavior,   i.e. a cyclic RG flow.      If the period of the RG flow is $\lambda$,   then the RG flow of the couplings implies 
\beq
\label{glambda} 
g(\ell + \lambda) = g(\ell)
\eeq
where $\ell = \log L $ is the logarithm of the {\it length}  scale $L$,
 such that increasing $\ell$ corresponds to a flow to low energies. 
 This can be interpreted as a discrete rather than continuous conformal symmetry.    
The RG period $\lambda$ is the fundamental parameter of such a flow,  and should be an RG invariant,  as it will be in the models defined in this article.       
Wilson  then  suggested a signature of such a flow is a periodicity of the S-matrix as a function of the  center of mass energy $E_{\rm cm}$:
\beq
\label{SmatrixPeriod}
S(e^{-\lambda} E_{\rm cm} , g ) = S( E_{\rm cm} , g ). 
\eeq
In Wilson's last work  with Glazek \cite{GW1,GW2},  motivated by results in nuclear physics 
\cite{nuclear},\footnote{The latter corresponds to a limit cycle in the IR.} 
 they  considered a quantum mechanical model in $D=0+1$ spacetime dimensions where a related signature of a cyclic RG is an infinite sequence of eigenstates 
with scaling behavior 
\beq
\label{ERussianDoll}
E_{n+1}  \approx  e^{\lambda} \, E_n  ~~~~~ n=1,2, 3, \ldots ~~~~ {\rm for ~ n ~ large}.
\eeq
A picturesque description of this behavior are nested  ``Russian Dolls",  i.e. Matryoshka Dolls,     wherein the spectrum  repeats itself  indefinitely in every cycle $\lambda$ as one probes higher energies, again at least up to some cut-off.      In relativistic models,    a Russian Doll RG is such that as one probes ever smaller length scales,   one finds the main structures repeat themselves up to some UV cutoff.   
For a general discussion of signatures of RG limit cycles we refer to the  work \cite{Elliptic}.\footnote{The ``Russian Doll" terminology was coined in \cite{LeClairSierra}. For readers not familiar with this ancient artisan craft,   we include an image in Figure \ref{Dolls}.}  
For a review of cyclic RG flows mainly aimed at applications to nuclear physics,   see 
\cite{BraatenReview}.

Concurrently with Glazek-Wilson's work,   with D.  Bernard \cite{BLflow} we independently  proposed cyclic RG flows for current-current perturbations in relativistic  $2D$ quantum field theory based on the  higher order beta-functions  proposed in \cite{GLM}.      Subsequently,    a many body version of Glazek-Wilson's model was proposed in \cite{LeClairSierraBCS} which is a generalization of BCS  superconductivity with an additional 
coupling which breaks time-reversal symmetry $\CT$,   and also exhibits cyclic RG behavior with the above Russian Doll property,  namely there is an infinite sequence of BCS condensates with the behavior \eqref{ERussianDoll}.    For the current-current perturbations considered in \cite{BLflow},    a relativistic  S-matrix was presented in \cite{LeClairSierra},  referred to as the cyclic sine-Gordon model below,   which had the anticipated infinite sequence of resonances with the Russian Doll behavior \eqref{ERussianDoll}.    Furthermore,  the resulting thermodynamic Bethe ansatz was studied indicating 
oscillations of the thermodynamic c-function in the UV \cite{Roman}.

Based on the above examples,   RG limit cycles can arise from marginal perturbations of conformal field theories.      In $D=2$,   theories with Lie group $G$ symmetry have many possible marginal perturbations based on 
$J \bar{J}$ operators where $J, \bar{J}$ are dimension $1$ operators corresponding the left,right  moving conserved $G$ currents.    
For $D=4$, marginal perturbations are much more limited.      Since fermions have dimension $3/2$,  one is led to consider $\phi^4$ perturbations,  since scalar fields $\phi$ have dimension 1.      However  RG flows for conventional perturbations like $g \phi^4$ have very limited properties based on the beta function $\beta_g \propto g^2$.     For  $g<0$ these are marginally irrelevant and there is no ultra-violet (UV) fixed point.     In the Higgs sector of the Standard Model of particle physics,   this is,  roughly speaking,   the origin of the so-called hierarchy problem,   namely,    if there is no UV fixed point,   the
natural scale for the Higgs boson should be much higher.   

The original motivation for  this article was to define new kinds of marginal perturbations for scalar fields   in 4 spacetime dimensions with  more interesting RG
flows than conventional $\phi^4$ interactions.         This led us to consider marginal operators with an interesting operator product expansion (OPE),   which parallels current algebra OPE's in 2D.    The models,    as defined below, are pseudo-hermitian: 
\beq
\label{pseudohermitian} 
 H^\dagger = \K H \K^\dagger ,  ~~~~~ \K^\dagger \K  =1, ~~~~~ \K = \K^\dagger,     ~~~~ \Longrightarrow ~~~ \K^2 = 1,
 \eeq   
 such that $H^\dagger$ is unitarily equivalent to $H$.   
Hamiltonians with this property were considered long ago by Pauli \cite{Pauli}.     
More recently,  pseudo-hermitian quantum mechanics has been developed in detail by  Mostafazadeh and others \cite{Mosta1,Mosta2,Mosta3},   and is by now a well-established and well  cited framework.\footnote{    
Pseudo-hermiticity,  as defined by \eqref{pseudohermitian} is not the same as $\CP \CT$ symmetric quantum mechanics,   which also has real eigenvalues if the $\CP \CT$ symmetry is unbroken \cite{BenderReview}.}    
Theories based on pseudo-hermitian hamiltonians have real eigenvalues but are nevertheless non-unitary in a strict sense.    However there exists various procedures for
defining unitary quantum mechanics from them,  which are model-dependent \cite{Mosta1}.    
The non-unitarity is  manifested in negative norm states,  as for the non-unitary minimal models of CFT \cite{BPZ} classified on the basis of the Virasoro algebra.    In fact,  there are many more non-unitary theories than unitary ones in this classification.   Projection onto unitary sub-Hilbert spaces is a common feature of 2D CFT,   for instance for Coulomb gas methods 
\cite{DotsenkoFateev},   which can be described as a BRST procedure \cite{BernardFelder}.

Let us mention that 
in \cite{38fold},   non-hermitian hamiltonians  with the additional structure \eqref{pseudohermitian} were classified depending on their discrete symmetries of charge-conjugation (particle-hole symmetry) 
$\CC$,   time reversal $\CT$ and parity $\CP$,   extending the classifications of Dyson-Wigner \cite{Dyson} and Altland-Zirnbauer \cite{AltlandZirnbauer}.   In this way the original 3-fold way of Dyson-Wigner grew to 10-fold  then to the  $38$ classes of hamiltonians found
in \cite{38fold},  where pseudo-hermiticity \eqref{pseudohermitian}  was incorporated as a constraint.        
The primary  application of these 38 universality classes is to open quantum systems which are non-unitary.    
Whether the Universe is an open or closed quantum system is debatable.    For instance in the search for a de Sitter/CFT correspondence,   it has been proposed that the 
CFT is non-unitary \cite{Strominger} and symplectic fermions \cite{LeClairNeubert} were considered in that context.     As we will see,  the pseudo-hermiticity of 
symplectic fermions strongly parallels that for the bosonic models in this paper.       
  Application of pseudo-hermitian hamiltonians
satisfying \eqref{pseudohermitian}   to  closed quantum systems  require additional selection rules to project onto unitary regimes,  and we will describe such a low energy regime for our model below.    More generally,   in principle a theory may involve  perturbations by irrelevant pseudo-hermitian operators such that the low energy  effective theory is
unitary.   There actually exists RG flows between minimal models of CFT in 2 spacetime dimensions where the UV theory is non-unitary whereas the IR fixed point is unitary.

Based on our 1-loop calculation of the RG beta functions below,    our  2-coupling model  corresponding to SU(2) broken to 
U(1)  has  a rich structure of RG flows.    Depending on the regime of couplings,    there is a line of UV and IR fixed points and also an unavoidable regime with a cyclic RG flow.\footnote{{\bf Note added:}   In subsequent work \cite{RDolls2},    we extended our calculation of the beta functions to 3-loops and showed that the 1-loop results of this paper are not spoiled to higher orders.}    
Cyclic  RG  flows have previously been considered as either impossible,  rare,  or  just curiosities,   because of various c-theorems and other considerations \cite{ctheorem,CardyCtheorem, Osborn,  JackOsborn,Schwimmer,Komargodski,Polchinski}.       The latter studies often utilize quantum fields coupled to gravity since the focus is on the conformal,  or trace anomaly of the stress-energy tensor.         Our model certainly exists by construction,  and as we showed in subsequent work \cite{RDolls2},  it  has a regime with  massless flows between  non-trivial conformal fixed points.    
Although our models  are non-unitary,    these massless flows still fit in the paradigm that all flows begin and end at conformal fixed points.   
  One should mention that in recent studies of massless flows between non-unitary minimal CFT's in 2D,   such flows are still consistent with a c-theorem \cite{Ravanini,MussardoLG,Tanaka,KlebanovMM,Negro},  where here $c=c_{\rm eff}$ is the 
  ``thermal" central charge that appears in the thermodynamic Bethe ansatz.    
  Furthermore,  there are examples of non-unitary minimal models that flow to unitary ones.        Returning to our model,    if  one is forced to  understand  the flows  for all regimes of the 2 couplings,   then a cyclic regime  is  inescapable.\footnote{{\bf Note added:}  This was studied to higher orders in \cite{RDolls2}.}   
Concerning the existence of these cyclic RG flows,     one needs to understand how our model 
circumvents certain,  perhaps hidden assumptions behind studies that would seem to rule out cyclic RG flows.  
One obvious answer is that non-unitary theories,  in particular those based on pseudo-hermitian hamiltonians,  were not considered before.     In \cite{Polchinski} for instance,  which relies heavily on the dilaton trick of  Komargodski and Schwimmer \cite{Schwimmer,Komargodski},    the assumption of unitarity is explicitly stated.     
 In addition to  this,  and we believe this is more important,    in order to formulate something like a c-theorem,    one needs   a well-defined  perturbation theory about both the UV and IR  fixed points  such that it applies to flows between fixed points,  and this was also assumed in \cite{Polchinski}.        For the cyclic regime of our model,   there are no such fixed points about which to perturb.         
One should  also mention that cyclic RG flows were found in a narrow,  non-unitary  region of couplings
for $\phi^6$ theory in $D=3 - \epsilon$ dimensions where $\phi$  is a matrix of scalar
fields \cite{Klebanov}.     The limit cycles in \cite{Grinstein} appear  different since it was proposed that they are actually conformally invariant,   rather than invariant under a discrete  conformal transformation.

 The  new type of marginal operators  defined below have a Lie algebraic structure in their OPE that is very similar to current-current perturbations in 2D.    
Although our model readily  generalizes to other Lie groups which would be interesting to develop,     we focus on Higgs-like fields with SU(2) symmetry for obvious reasons.   
 Our main result 
 is  that such marginal operators have a very rich structure of RG flows,   including both fixed points and flows between them,  and also a regime that  exhibits cyclic RG behavior.
 In order to shed light on what are the consequences of the cyclic RG in our 4D  models,  in Section V we review how this cyclic RG  behavior is manifested as an infinite  spectrum of resonances  with Russian Doll behavior 
 for 2D  integrable QFT's with beta functions that are identical to those for our $4D$ models. 
 
Since we study Higgs-like models,    it would be remiss not to 
   study spontaneous symmetry breaking (SSB) in our model,  and we do so  in Section VI.   We  show that there are an infinite number of vacuum expectation values 
 satisfying the Russian Doll scaling.   These are analogous to the Russian Doll BCS condensates found in 
 \cite{LeClairSierraBCS}.      The interplay between the Higgs mechanism and a cyclic RG flow is clearly something worth investigating  and a complete understanding is beyond the scope of this article.     Nevertheless,   based on our analysis of spontaneous symmetry breaking,   we  speculate that a cyclic RG  could perhaps resolve the hierarchy problem and is even perhaps at the origin of ``families",  otherwise called  ``generations",   in the Standard Model.     The idea is that each family  is one of an infinite number of Russian Dolls with nearly identical properties apart from the exponentially increasing spectrum of masses as in 
 \eqref{ERussianDoll}.        

\section{Definition of the models}

\def\Phidagger{\Phi^\dagger}

\subsection{The free theory and its $\CC, \CP, \CT$ symmetries}

\def\htilde{\tilde{h}}

Throughout this article,  we consider relativistic quantum fields in $D=d+1=4$ spacetime dimensions,  apart from the brief review of some results 
in $D=2$  in Section V.      
Introduce two  independent  doublets of  complex  bosonic  spin-0 fields,  
$\Phi_{i}$ and $\Phitilde_{i}$, ${i =1,2}$
and their hermitian conjugates $\Phi^\dagger$, $\Phitilde^\dagger$.     
Under SU(2) transformations,   $\Phi (x) \to U \Phi (x)$,   where $U$ is a $2\times 2$  unitary SU(2) group matrix acting on the ``i" indices in the above equation and the same for $\Phitilde$.     The symmetry is actually U(2) = SU(2) $\otimes$ U(1) 
where the extra U(1) is a hypercharge.  
 These fields  transform as  two Higgs doublets.     
The free action is the standard one for free  massless bosonic fields.  
\beq
\label{Sfree}
\CS_{0} = \int d^4 x \( \d_\mu \Phi^\dagger \d^\mu \Phi  +  \d_\mu \Phitilde^\dagger \d^\mu \Phitilde \),
\eeq
where $\Phidagger \Phi = \sum_{i=1,2} \Phidagger_i \Phi_i $ etc.    The free theory has global $U(2)\otimes U(2)$ symmetry since the fields
$\Phi,\Phitilde$ are independent,   however the interactions introduced below will  break this down to the diagonal  $U(2)$,   where $\Phi$ and $\Phitilde$ transform in the same way under this diagonal  $U(2)$.

Let us review the quantization of this theory in order to point out certain aspects of its discrete symmetries we will need to refer to \cite{Peskin}. 
For this purpose we can suppress the $i$-indices in $\Phi_i$,   namely let us first deal with a single ``$i$" component of $\Phi_{i=1,2}$.    
Expand  the field in terms of particle creation/annihilation operators $a_\pm,  a^\dagger_\pm$:
\barray
\label{as}
\nonumber 
\Phi (x) &=&  \int \frac{d^3 \kvec}{( 2 \pi)^{3/2}} \inv{\sqrt{2 \omega_\kvec}} \( a^\dagger_- (\kvec ) e^{- i k\cdot x}  + a_+(\kvec) e^{i k\cdot x } \) 
\\ 
\Phi^\dagger  (x) &=&  \int \frac{d^3 \kvec}{( 2 \pi)^{3/2}} \inv{\sqrt{2 \omega_\kvec}} \( a_-(\kvec) e^{i k\cdot x}  + a^\dagger_+ (\kvec) e^{-i k\cdot x } \), 
\earray 
where 
$ k \cdot x = \omega_\kvec t - \kvec \cdot \xvec $ with $\omega_\kvec = |\kvec| $.  
Canonical quantization of the boson field yields 
\beq
\label{canonical}
[a_\pm (\kvec), a^\dagger_\pm (\kvec')] = \delta^{(3)}  (\kvec - \kvec' ),
\eeq
and the hamiltonian is
\beq
\label{freeH}
H_{0}  = \int  d^3 \kvec  \, \, \omega_{\kvec} \(   a^\dagger_+ (\kvec) \, a_+(\kvec) + a^\dagger_-(\kvec) \, a_-(\kvec)  \). 
\eeq

The above free theory is necessarily  invariant under $\CC\CP\CT$,   which we now review in order to make certain points below.
Consider first charge conjugation implemented by the operator $\CC$:   
\beq
\label{Cconj2}
\CC \CC^\dagger = 1, ~~~ \CC = \CC^\dagger, ~~~ \CC^2 =1. 
\eeq 
 On the $a, a^\dagger$  operators it acts as 
 \beq
\label{ChargeC}
\CC a_\pm(\kvec) \CC = a_\mp (\kvec), ~~~\CC a^\dagger_\pm (\kvec) \CC = a^\dagger_\mp (\kvec).
\eeq
This implies 
\beq
\CC \Phi (x) \CC = \Phi^\dagger (x)\, ~~~~ \Longrightarrow ~~~~~~\CC H_0 \CC = H_0= H^\dagger_0. 
\eeq
Based on  $\CC$  we  identify  particles as those created by $a^\dagger_+$ and anti-particles as created by $a^\dagger_-$. 
This operator $\CC$  will play a central role below since it will be broken by interactions.

Let us also review the standard  parity  $\CP$ and time-reversal symmetry $\CT$. 
We take them to also satisfy constraints such as in \eqref{Cconj2}.    
Since parity flips the sign of momentum: 
\beq
\label{Paritydef}
\CP  a_\pm (\kvec) \CP =  a_\pm (-\kvec),  
~~~\CP  a^\dagger_\pm(\kvec) \CP =  a^\dagger_\pm (-\kvec), ~~~~~~\Longrightarrow ~~~~\CP \Phi(t,\xvec) \CP = \Phi(t, - \xvec), ~~~
\CP H_0 \CP = H_0.
\eeq
Time-reversal also flips the sign of $\kvec$ and only differs from $\CP$  in that it is anti-unitary, namely $\CT z \CT = z^*$ for $z$ a complex number.   This leads to 
\beq
\label{TimeRev} 
\CT  a_\pm(\kvec) \CT = a_\pm (-\kvec), ~~~\CT  a^\dagger_\pm  (\kvec)\CT = a^\dagger_\pm (-\kvec)~~~~
\Longrightarrow ~~~ \CT \Phi (t, \xvec) \CT = \Phi (-t, \xvec ), ~~~~~ \CT H_0 \CT = H_0.
\eeq
This implies 
\beq
\label{PT}
\CP\CT  a_\pm(\kvec) \CP \CT = a_\pm (\kvec) ,
\eeq
and
\beq
\label{PT2}
\CP\CT  H_0  \CP \CT = H_0.
\eeq

The full  Hilbert space $\CH$,   which diagonalizes the free hamiltonian $H_0$,  consists of multi-particle  states $|\psi\rangle$ of the form 
\beq
\label{states}
\CH = \{ |\psi \rangle \} , ~~~ | \psi \rangle =  |(\kvec_1, s_1), (\kvec_2, s_2), \ldots  (\kvec_n,  s_n)  \rangle \equiv  a^\dagger_{s_1} (\kvec_1 )  a^\dagger_{s_2} (\kvec_2 ) \cdots a^\dagger_{s_n} (\kvec_n ) |0\rangle  ~~~~~
s_i \in \{\pm \}.
\eeq
In $\CH$ all states have a positive norm such that 
\beq
\label{freenorm}
\langle \psi' |\psi\rangle =  \delta_{\smallpsi'  \smallpsi}  >0,   ~~ \Longrightarrow  ~~~   \sum_\smallpsi   |\psi\rangle \langle \psi | = 1,
\eeq
where $\delta_{\smallpsi' \smallpsi}$  symbolically contains overall energy-momentum conserving  Dirac delta-functions.

\def\daggerk{\daggerc}

\subsection{Pseudo-hermitian marginal perturbations}

As discussed in the Introduction,  the original  goal of this paper was  to find marginal perturbations with potentially interesting patterns of RG flows 
for quartic interactions of scalar fields.  
The natural operators,   $(\Phidagger \Phi )^2$  are important  parts of the  standard Higgs potential for $\Phi$,   however the RG flows are not very fertile and  the theories are generally not well-defined in the UV.     In fact,  this is essentially the origin of the so-called hierarchy problem.
As we now describe,   there is a different construction of marginal  operators which can have a much richer  pattern of RG flows as a consequence of a Lie-algebraic structure of their operator product expansion.          
This will require that the  resulting hamiltonian be pseudo-hermitian \eqref{pseudohermitian}.   To motivate this construction,    note that 
there is another discrete symmetry of the free hamiltonian in the form \eqref{freeH}:
\barray
\label{Kdef}
&~& 
\K a_\pm (\kvec)  \K = \pm a_\pm  (\kvec) ,  ~~~ \K a^\dagger_\pm (\kvec) \K  = \pm a^\dagger_\pm (\kvec) , \\
&~& ~~~~~~\Longrightarrow ~~~\K H_0 \K = H_0 = H^\dagger_0.
\earray
Every state $\psi$ in the Hilbert space defined above,  spanned by states of the form \eqref{states},  has  a well defined $\K$ eigenvalue
\beq
\label{Kpsi}
\K |\psi\rangle = \K_\smallpsi  \, |\psi\rangle, ~~~~~  \K_\smallpsi \in \{ \pm 1\}, ~~~~~ \K |0\rangle = |0\rangle.
\eeq
 
The operator $\K$ does not act locally on the fields $\Phi$,  unlike $\CC, \CP, \CT$.    
The operator $\K$ can be expressed in terms of the $U(1)$  charge $\qop$ operator.  
Let $N_\pm$ the number operators for particles created by $a^\dagger_\pm$:
\beq
\label{Nops}
N_\pm = \int d^3 \kvec \, a^\dagger_\pm (\kvec)  a_\pm (\kvec).
\eeq
Then the U(1) charge is 
\beq
\label{qop}
\qop = N_+ - N_-  
\eeq
and one has 
\beq
\label{charges}
 [\qop,  a_\pm] = \mp a_\pm , ~~~~[\qop,  a^\dagger_\pm] = \pm a^\dagger_\pm. 
 \eeq
One can then identify 
\beq
\label{KU1}
\K = e^{i \vartheta \Nanti} ~~ {\rm with} ~  \vartheta = \pi ~ {\rm and} ~  \Nanti \equiv N_- ~~~~ \Longrightarrow~~ [\qop,  \K] = 0,
\eeq
where $\Nanti$ counts the number of anti-particles.     
Clearly,   $\K^\dagger K = 1$  and  $\K^2 = 1$  since  $\Nanti$ is an integer on the Hilbert space.  
 The $\K$ operator does not commute with charge conjugation $\CC$.   One has
 \beq
 \label{CK}
 \CC \K = \K \CC \, e^{i \pi \qop},
 \eeq
 where we have also used that  $\qop$ is an integer on the Hilbert space,
 and 
 \beq
 \label{CQ}
 e^{i \pi \qop} \CC = \CC e^{i \pi \qop}  .
 \eeq
 On 1-particle states where $\qop = \pm 1$,  eq.  \eqref{CK} implies that $\K$ anti-commutes with $\CC$.

For any operator $A$,   define it's  $\K$-hermitian conjugate $A^\daggerk$ as follows
\beq
\label{Aconj}
A^\daggerk = \K A^\dagger \K . 
\eeq
In particular,   we will need $\Phi^\daggerk = \K \Phi^\dagger \K$.   
Unlike $\CC, \CP, \CT$,   the operator $\K$ acts non-locally on the fields,  thus $\Phi^\daggerk$ should be viewed as a non-local field.  
The operator $ \Phi^\daggerc \Phi$ is not its own hermitian conjugate,   but is invariant under $\daggerk$ conjugation:
\beq
\label{biConj}
( \Phi^\daggerc \Phi)^\daggerk  = \Phi^\daggerc \Phi .
\eeq
Based on this,   for a hermitian $2\times 2$ matrix $\tau^a$ define the operator 
\beq
\label{J}
J^a  \equiv  \Phi^\daggerc  \tau^a \Phi = \sum_{i,j}  \Phi^\daggerc_i  \tau^{a}_{ij} \Phi_j ~~~~~~ (\tau^a)^\dagger = \tau^a,
\eeq
and similarly for $\Jtilde^a$.   Our  reason for defining the operators $J^a$  is the interesting OPE such operators have,   as described in Section IV,   which leads to simple algebraic formulas for their beta-functions,  which can be extended to other Lie groups.

We can now define our models.    Define the marginal operators 
\beq
\label{OAdef} 
\CO^A (x) = \sum_{a,b} d^A_{ab} \,  J^a (x) \Jtilde^b (x) 
\eeq
for some coefficients $d^A_{ab}$.     The models are then defined by the action 
\beq
\label{perturbations}
\CS = \CS_{0}  + 4 \pi^2  \int d^4 x \,  \sum_A g_A  \CO^A (x). 
\eeq
The factor of $4 \pi^2$ is introduced to simplify the beta-functions of the next section.   
The operators $J^a$ are not hermitian but pseudo-hermitian:
\beq
\label{Jdag} 
J^\daggerc = J ,   ~~~~~     \Jtilde^\daggerc = \Jtilde.
\eeq
Thus the  interacting hamiltonian is pseudo-hermitian:
\beq
\label{Hdag2}
H^\dagger = \K H \K , ~~~ \Longrightarrow ~~ H^\daggerc = H .
\eeq
In the next section we will consider   $\K$-conjugation  $\daggerc$ as insertions of the operator $\K$ in correlation functions.

 \subsection{$\CC, \CP$ and $ \CT$ for the interacting theory:   breaking of  $\CC \CP$}

The free theory with hamiltonian $H_0$ is invariant under separate $\CC, \CP, \CT$ symmetries thus respects the $\CC\CP\CT$  theorem since  it does not break the combination 
$\CC\CP\CT$.      Since the proof of the $\CC\CP\CT$  theorem assumes hermiticity and unitarity,     $\CC\CP\CT$  is actually  broken for the interacting theory.      Since 
$\CC$ exchanges 
$a^\dagger_\pm$ according to \eqref{ChargeC},   and $\K$ distinguishes between particles and anti-particles,    then it is clear that $\CC$ is broken.   
This is manifested in the fact that $\CC$ does not commute with $\K$,    as \eqref{CK} shows.   On the other hand,  based on \eqref{PT},   $\K$ does commute with
$\CP\CT$,  so the breaking of CPT is due only to the breaking of $\CC$.      Furthermore,   since $\K$ commutes with both $\CP$ and $\CT$ separately,   then 
$\CP$ is unbroken.     Thus our model breaks $\CC\CP$.    

More generally,  if  $\CC\CP\CT$  is broken,   then this could allow  violations of the spin-statistics theorem since the 
$\CC\CP\CT$ theorem assumes a hermitian hamiltonian and unitarity.         This is not relevant for this article where we quantized the scalar field
$\Phi$ as a boson.       However the  pseudo-hermitian properties of our model have very strong parallels with that
of symplectic fermions,   where a scalar field can be consistently quantized as a fermion \cite{LeClairNeubert,CYLeeSpinStat}.\footnote{It is an interesting exercise to
specialize the 38-fold classification in \cite{38fold} to relativistic QFT and we hope to publish such results in the future.}

\section{Addressing non-unitarity:   negative norm states and a low energy  unitary regime}

Our models could have applications to statistical mechanics in 4 spatial dimensions where the positivity constraints of unitarity 
are not so stringent.     For potential applications to  relativistic particle  physics in $3+1$ spacetime dimensions,     a consistent unitary quantum mechanics would be desirable,  if not in  fact necessary,   and this is addressed in  this section.    
The subject of Quantum Mechanics for pseudo-hermitian hamiltonians is by now well developed,    with extensive reviews 
\cite{Mosta1,Mosta2,Mosta3}.   These works show that in principle one can still define a unitary theory from a pseudo-hermitian hamiltonian \cite{Mosta1},  however  the procedure for doing so is model dependent and we could not apply it directly to our theory.   
Thus  in  this section we study this issue in detail.      This will lead us to conclude that although our model is  strictly speaking non-unitary,
as we will explain, it is in fact  perfectly unitary at low enough energies,   specifically below the energy  threshold for pair-production.   

To set the stage,  let us first review the  Born rule for probabilities in standard unitary quantum mechanics in the Schr\"odinger picture.    
Assume we are given a Hilbert space $\CH$ with a conventional positive definite inner product.   
Namely for  $|\psi \rangle \in \CH$,     there exists an inner product such that $\langle \psi' | \psi \rangle$ is positive definite.   Indeed,  the  states in \eqref{states} comprise such a Hilbert space.       
At time $t=t_0 =0$ the states need to be properly normalized such that $\langle \psi | \psi \rangle = 1$.      
There exists a basis in $\CH$,     $|\psi_n\rangle \in \CH$ where for simplicity of notation we assume $n$ is discrete,   
$n = 1, 2, \ldots \infty$.     Then 
\beq
\label{psins}
|\psi \rangle = |\psi (t=0) \rangle = \sum_n  c_n \,  |\psi_n \rangle,      ~~~~\langle \psi_m | \psi_n \rangle = \delta_{nm} ,
\eeq
for some complex numbers $c_n$.     
Then 
\beq
\label{norm1} 
\langle \psi | \psi \rangle = 1  ~~~~~~\Longrightarrow ~~~ \sum_n  c_n^* c_n = 1 ,
\eeq
where $c_n^*$ denotes complex conjugation.    According to the Born rule,   $|c_n|^2$  represents the probability that $|\psi \rangle$ is measured to be in the state $|\psi_n\rangle$.     Given a hamiltonian $H$,  or any other operator  $A$ on $\CH$,   then $A^\dagger$ is defined in the standard way with respect to the above inner product.     
Let us now turn to pseudo-hermitian hamiltonians in general.

\subsection{General properties of pseudo-hermitian Quantum Mechanics}
 
Pseudo-hermitian hamiltonians have many desirable properties that parallel those in ordinary hermitian quantum mechanics,  and we here list the most important
for such a theory generally,  i.e. properties that are independent of the model.

\bigskip

\noindent (i) ~~{\it Modified indefinite  metric  on the Hilbert space}
\medskip

Suppose the hamiltonian operator $H$ is pseudo-hermitian with the structure in \eqref{pseudohermitian},  which we repeat here:
\beq
\label{HdaggerAgain}
H^\dagger = \K \, H \, \K^\dagger,    ~~~ \K^\dagger \K =1, ~~~~~ \K^\dagger = \K, ~~~~ \Longrightarrow~~~ \K^2 =1 .
\eeq
In other words,   $H^\dagger$ is unitarily equivalent to $H$.  
This leads us to define a new inner product which includes an insertion of the operator $\CK$:
\beq
\label{Knorm}
\brabra \psi' | \psi \ketket \equiv \langle \psi' | \CK | \psi \rangle = \K_\smallpsi \, \delta_{\smallpsi'\smallpsi}
\eeq
In this new inner product,    kets have an accompanying $\K$ but bras do not.    
In general,    this new inner product  can have  negative norm states $|\psi\rangle$ if $\K_\smallpsi$ is negative.   
This implies negative probabilities since \eqref{norm1} becomes 
\beq
\label{negprob}
 \sum_n  \K_n \, |c_n|^2 = 1.
 \eeq

\bigskip

\noindent (ii) ~~{\it  Pseudo-hermitian operators correspond to observables with real eigenvalues}
\medskip

For any operator $A$ on $\CH$,     define its pseudo-hermitian  conjugate as follows:
\beq
\label{pseudodag}
A^\daggerk \equiv \CK \, A^\dagger  \, \CK .
\eeq
Then the hamiltonian is pseudo-hermitian:
\beq
\label{pseudoH}
H^\daggerk = H .  
\eeq
$A^\daggerk$ is the proper conjugation based on the $\CK$-metric in \eqref{Knorm},   namely
\beq
\label{Astar}
\brabra \psi' | A | \psi \ketket^* =  \brabra \psi | A^\daggerk  | \psi' \ketket ,
\eeq
where $*$ denotes ordinary complex conjugation.  
We define  a pseudo-hermitian operator $A$ as one that satisfies $A^\daggerk = A$.     
From \eqref{Astar},   one concludes that pseudo-hermitian operators,   in particular the hamiltonian H,  has real eigenvalues. 
In other words,   $H$ is effectively hermitian but on a Hilbert space with indefinite norm.   
Any operator satisfying $A^\daggerk = A$ in principle can correspond to an observable.   
One can easily establish that this pseudo-hermitian adjoint satisfies the usual rules, e.g.\
\barray\label{rules1}
   (AB)^\daggerc &=& B^\daggerc A^\daggerc \,, \nonumber\\
   (a A + b B)^\daggerc &=& a^* A^\daggerc + b^* B^\daggerc \,,
\earray
where $A,B$ are operators and $a,b$ are complex numbers.

\bigskip

\noindent (iii) ~~{\it  Conservation of probability}
\medskip

 The hamiltonian determines the time evolution of  a state $|\psi \rangle =|\psi (t=0)\rangle$,    namely 
 $| \psi (t) \rangle =  \,e^{-i H t} |\psi \rangle$.       By conservation of probability we mean that the norm of states is preserved under time evolution.     More generally:
 \beq
 \label{timeevolution}
 \brabra \psi' (t) | \psi (t) \ketket = \langle \psi'| e^{i H^\dagger t } \, \K \, e^{-i H t} \, | \psi \rangle 
 = 
 \langle \psi'| \K e^{i H t }  \K^2  \, e^{-i H t} \, | \psi \rangle 
 = 
 \brabra \psi' (0) | \psi (0) \ketket 
 \eeq
 In general,   the modified $\K$-metric $\brabra \psi' | \psi \ketket$ introduces negative norm states which imply negative probabilities.  The above statement of unitarity just states that the sum of these probabilities is maintained under time evolution,
 but one must still deal with negative norm states due to the negative probailities.    
 
 \bigskip
 
 \noindent (iv) ~~{\it  Projection operators}
\medskip

The properties of $\K$ naturally lead to projection operators onto positive or negative norm states.      
 Since $\K^2 =1$,   states can be classified according to $\K = \pm 1$.     Namely,
 \beq
 \label{Kvalues}
 \K | \psi_n \rangle = \K_n  \, | \psi_n \rangle,   ~~~~  \K_n \in \{\pm 1 \}
 \eeq
 Define the operators\footnote{We adopt the notation $\pmdot$ to refer to $\K=\pm$ norm states in order to avoid confusion with $\pm$ in the definition of 
 the particle creation operators $a^\dagger_\pm$.} 
 \beq
 \label{Projectors}
 \Ppm =  \frac{1 \pm \K}{2}.
 \eeq
 They satisfy the usual algebra of projection operators:
 \beq
 \label{Projectors2}
 \Pplus  + \Pminus =1, ~~~~~ \Ppm^2 = \Ppm  , ~~~~~  \Pplus\Pminus = 0.
 \eeq
 In addition,  one has 
 \beq
 \label{ProjectorsK}
 \K \Ppm = \Ppm \K = \pm \Ppm.
 \eeq
 Thus the Hilbert space decomposes as 
 \beq
 \label{Projectors3} 
 \CH = \Hplus  \oplus \Hminus , ~~~~~ | \psipm \rangle \in \Hpm , ~~~~~ \CK |\psipm \rangle = \pm |\psipm \rangle .
 \eeq
 One has 
 \beq
 \label{norms2}
 \brabra \psipm | \psipm \ketket = \langle \psi | \Ppm  \K \Ppm | \psi \rangle = \pm \langle \psipm | \psipm \rangle, 
 ~~~~~ 
  \brabra \psiminus  | \psiplus  \ketket  = \langle \psi | \Pplus \K \Pminus  | \psi \rangle =0 
  \eeq
  
 If we project onto the sub Hilbert space $\Hplus$,     then all states $|\psi_n \rangle \in \Hplus$ have positive norm,  i.e. 
 $\K_n =+1$,   which implies positive probabilities $|c_n|^2 \geq 0$.        This positivity is preserved under time evolution:
 \beq
 \label{projectHilbert}
 \brabra \psipm (t) | \psipm (t) \ketket = \brabra \psipm (0) | \psipm (0) \ketket , ~~~~~
  \brabra \psiminus  (t) | \psiplus  (t) \ketket =  \brabra \psiminus (0) | \psiplus (0) \ketket = 0.
  \eeq
However this is insufficient to define a physically consistent quantum mechanics since $\K$ does not commute with $H$,  otherwise 
$H^\dagger = H$,  which is not the case by construction.       This implies that if one prepares a state at time $t=0$ that is in $\CH_{\plusdot}$,  then negative norm states
will be generated under time evolution:
\beq
\label{generationofneg}
e^{-i H t}  |\psi_{\plusdot} \rangle ~ \notin~ \CH_{\plusdot} 
\eeq
In order to have a consistent quantum mechanics one needs an additional {\it selection rule}  of sorts such that 
$e^{-i H t}  |\psi_{\plusdot} \rangle ~ \in~ \CH_{\plusdot}$.    At this point,  if such a selection rule exists,  it doesn't follow generally from pseudo-hermiticity,  but rather
must be model dependent.       In our model,   the free hamiltonian $H_0$ does commute with $\K$,  but the interacting part does not.     It is thus natural to consider this issue
in the interaction picture,  in particular in scattering theory,  which we turn to next.

  \def\bra{\langle}
  \def\ket{\rangle}

 \subsection{Specialization to our model:   Scattering theory and a  kinematic unitary regime}
  
 In the interaction picture  the hamiltonian is separated  into a free part $H_0$ and the interaction:
\beq
\label{Hint}
H = H_0 + H_{\rm int} .
\eeq
For the purposes of this section,   we again only consider a single component $\Phi$ as in Section IIA,   and consider the pseudo-hermitian  interaction
$H_{\rm int} = \int d^3 x \, (\Phi^\daggerk \Phi)^2$. 
Since $H_{\rm int}$ is local,   then far from the interaction region  the asymptotic  states in the far past and far future are 
eigenstates of the free hamiltonian,  and one can separate out this time evolution with $H_0$  which leads to a well-defined perturbation theory based on 
$H_{\rm int}$.      For hermitian quantum mechanics this leads to the standard formalism of scattering theory,   where positive probabilities lead to physically well-defined
and meaningful cross sections.    

Let us adapt  this scattering formalism to pseudo-hermitian theories where the negative norm states cannot simply be projected out consistently.       
Define  the operator
\beq
\label{Omega}
\Omega (t) = e^{i H t} e^{-i H_0 t}  
\eeq
where as above  we have taken the reference time $t_0 =0$ for simplicity,  otherwise $t$ in the above equation should be replaced by $t-t_0$.          
Using $[ \K, H_0 ] = 0$,   one can easily show that 
\beq
\label{Omegadag} 
\Omega^\daggerk \Omega = 1.
\eeq
The so-called in and out states are defined by the M\o ller operators $\Omega_\pm$:
\beq
\label{inout}
|\psi \rangle_{\rm in} = \Omega_- |\psi \rangle, ~~~~|\psi \rangle_{\rm out } = \Omega_+ |\psi\rangle , ~~~~~\Omega_\pm = \lim_{t\to \pm \infty}  \Omega (t).
\eeq
Given  the $\K$-metric,   the S-matrix operator is now defined as follows:  
\beq
\label{Smatrix}  
_{\rm out} \brabra \psi' | \psi \ketket_{\rm in} \equiv  \langle \psi' | S | \psi \rangle  ~~ \Longrightarrow ~~  S = \Omega_+^\dagger \K \Omega_- = \K \, \Omega_+^\daggerk \Omega_- \, \, .
\eeq
Using $S^\daggerk = \Omega_-^\daggerk \Omega_+ \, \K $,  one finds 
\beq
\label{Sunitarity}
S^\daggerk S = 1.  
\eeq

The formula \eqref{Sunitarity} generalizes the usual statement of the unitarity of the S-matrix and  is equivalent to the fact that the norm of states is time independent,  i.e. \eqref{timeevolution}.   Thus by itself,  it does not resolve the  problem with negative norm states.   To address this,   one needs to consider physical quantities  that have a probabilistic meaning,  such as those in  the optical theorem.      To this end define the  usual $T$-operator  as 
\beq
\label{Ts}
S = \K + i T,   ~~~~~ S^\daggerk = \K - i T^\daggerk \,
\eeq
The $\K$ in the above equation comes from the $\K$ in the inner product if $T=0$.    The generalized unitarity relation \eqref{Sunitarity} expressed in terms of $T$ reads 
\beq
\label{optical}
i \( T^\daggerk  \K - \K T \) = T^\daggerk T
\eeq
Taking an inner product of the above $\langle \psi | \sim | \psi\rangle$ and inserting a complete set of states on the right hand side, $\K_\smallpsi$ cancels from both sides and one obtains  
\beq
\label{optical2} 
2  \, {\rm Im}  \( T_{\smallpsi \smallpsi} \) = \sum_{\psi'}   \K_{\smallpsi'} | T_{\smallpsi\smallpsi'} |^2 
\eeq
where we have defined  $T_{\smallpsi \smallpsi'} = \langle \psi | T | \psi'\rangle$.     The above is the generalization of the optical theorem for pseudo-hermitian theories.
If there are negative norm intermediate states $|\psi'\rangle$,  then they imply negative residues of kinematic poles in the S-matrix.   
It was first obtained by Cheng-Yang Lee \cite{CYLee} for scalar symplectic fermions \cite{LeClairNeubert},  where the precise nature of the pseudo-hermiticity is  essentially identical  to that of our bosonic model.     The above formula \eqref{optical2}  clearly shows the implications of  intermediate states with negative norm 
 $\K_{\smallpsi'} = -1$,  since for unitary theories all   $\K_{\smallpsi'} = +1$ and the right hand side is interpreted as the sum over positive transition amplitudes,  which leads to physically meaningful  cross sections.       

In order to have an effectively unitary theory with positive probabilities,  an  additional selection rule is needed such that {\it only} intermediate states with 
$\K_{\smallpsi'} = +1$ contribute to the right hand of \eqref{optical2}.    There is no such selection rule  in  our model without adding some additional properties not
already contained in its definition.      However there exists such a selection rule based simply on kinematics,  as we now explain.      
 Our model is a quantum field theory with the multi-particle Hilbert space \eqref{states},  
thus as usual we factor out the overall energy-momentum conserving delta-functions and define the amplitudes $\CM$:
\beq
\label{withdeltas}
T_{\smallpsi\smallpsi'} = (2 \pi)^4  \delta^{(4)} \( p_\smallpsi - p_{\smallpsi'}\) \CM_{\smallpsi \smallpsi'} \, .
\eeq
Consider adding a mass term to $H$ such that the particles have mass $m$.  We will do this below when we consider spontaneous symmetry breaking. 
Furthermore consider an initial state with only particles and  no anti-particles such that the in-state has positive norm,  and to further simplify matters let the in-state have only 
2 particles each of positive norm designated simply as $|++\rangle \equiv a^\dagger_+ a^\dagger_+ |0\rangle $.   Then the lowest energy intermediate state $|\psi'\rangle$ with negative norm 
that is consistent with the $U(1)$ charge $\qop$ conservation 
is $|+++- \rangle$ which requires particle/anti-particle  pair production $|+-\rangle$.   
We have used the fact that for a charge neutral operator $A$,   since $ [\qop, H] =0$,   one has 
\beq
\label{Qconservation}
 \brabra \psi' | A |\psi \ketket =0 ~~{\rm if} ~ \qop_{\smallpsi'} \neq \qop_\smallpsi .
 \eeq
If the total incoming energy is below the 4-particle threshold,  namely $s< (4m)^2$ where $s= (p_1 + p_2)^2$ is a  Mandelstam variable,   
then this process is kinematically forbidden.     Thus if one considers only 2 body scattering,   namely $2 \to 2$ particles,    then this process is unitary.   
The right hand side of \eqref{optical2} has only positive probability contributions which leads to standard formulas for the total cross section \cite{Peskin}:     
\beq
\label{optical3}
2 \, {\rm Im} \CM \( \kvec_1 , \kvec_2 \to \kvec_1 , \kvec_2 \) = 2 E_{\rm cm} p_{\rm cm} \, \sigma_{\rm tot} \( \kvec_1, \kvec_2 \to {\rm anything} \).
\eeq

In summary,  for 2-body elastic scattering of particles,  if the total energy is below the threshold for pair production,  the theory is effectively unitary.  
This result is analogous to the  low energy scattering of electrons $e_-$ below the threshold for $e_+ e_-$ pair production,  commonly referred to 
as  ``electron–electron M\o ller  elastic scattering"  $e_- e_- \to e_- e_-$, 
which is known to be unitary if one ignores production of  electron/positron pairs.\footnote{In full quantum electro-dynamics (QED) 2 body scattering can also radiate  photons,
but this amounts to almost negligible effects.   In any case our model consists of only particles and anti-particles since we have not gauged the U(1).} 
 There is a  difference between  our model and QED  since  both  electrons and positrons  have positive norm,  nevertheless the kinematic constraint still holds
 and the formula \eqref{optical} applies to both.   In the non-relativistic limit where all scattering is elastic,   namely the number of incoming and outgoing particles is the same, 
 our model is unitary.   

A similar,  but essentially different,  selection rule that restores unitarity is to consider the model at finite density by introducing a chemical potential $\mu$:
\beq
\label{chemicalpotential}
H \to H - \mu \qop = H-\mu \(N_+ - N_- \).
\eeq
For positive $\mu$,   it costs extra energy to add an anti-particle with $N_- = \Nbar > 0$ whereas adding particles $N_+$ is energetically cheap.   Thus with high enough
chemical potential,  production  of anti-particles  with negative norm is suppressed.  

The above considerations  open up the possibility of applying the  pseudo-hermitian bosonic models of this paper,   and also symplectic fermions,  to 
condensed matter physics,    where energies are low enough,  in fact typically non-relativistic,   which are effectively unitary since one can ignore pair production at these 
energies.   Symplectic fermions have a Fermi surface,  thus the above discussion of $\CC$-conjugation corresponds to particle-hole symmetry.     
Some such applications were discussed for symplectic fermions in \cite{LeClairNeubert}.    Let us also mention that 
since the groups Sp(4) = SO(5),    the N=2 component symplectic fermion has order parameters for anti-ferromagnetism,   superconductivity,  and also a potential pseudo-gap,
and can be studied as a toy model of high $T_c$ superconductivity \cite{KapitLeClair}  where the role of SO(5) was first proposed by S.-C. Zhang \cite{SCZhang}.\footnote{ In light of the above remarks,  the  main shortcoming  of the work \cite{KapitLeClair}   is not  necessarily  that it is non-unitary,   but rather  that it is formulated in the continuum without a lattice,  whereas the lattice  is known to play a fundamental role in obtaining a d-wave  superconducting gap.   As a toy model of superconductivity,  it still has a Fermi surface,  and  it is distinguished by the fact the 4-fermion interactions closely parallel the interactions in the Hubbard model,  and since they are marginal,    they imply a fermionic critical point in 2+1 dimensions that can be reached
in $4-\epsilon$ dimensions.   This critical point was studied to 2-loops in \cite{LeClairNeubert}.} 
A related point is that one can in some cases efficiently obtain novel results for non-relativistic models such as the Lieb-Liniger gas by taking a non-relativistic limit of 
an appropriate relativistic one,  in this case the sinh-Gordon model \cite{MussardoLieb}.


\section{Operator product expansion and the renormalization group}

In order to study the Renormalization Group,   instead of working with Feynman diagrams that implement the operator $\K$ insertions,
it is simpler to derive the main results using  the operator product expansion (OPE) of the marginal operator perturbations. 
In this section we present the results we need  to one loop using OPE's.   Some of the formulas below are borrowed from 
our study of symplectic fermions in \cite{LeClairNeubert} which are also pseudo-hermitian,   and it was shown there that an OPE approach agrees with Feynman diagram perturbation theory at least to 1-loop.          In fact,   some aspects of our construction
can be thought of as a  bosonic version of symplectic fermions,  but not the obvious one which is hermitian where $\daggerc$ is just $\dagger$. 
     Recently,   general properties of correlation functions for
pseudo-hermitian  theories were  studied in \cite{Chengdu} based more on the path integral and Feynman diagrams.

\subsection{RG from the operator product expansion}

Based on  the action \eqref{perturbations},  
consider  the correlation function
$\langle \X\rangle$ for arbitrary $\X$ to second order in $g_A$.  
Assume the closed OPE algebra 
\beq
\label{OOPE}
\CO^A (x) \CO^B (0) \sim    \inv{4 \pi^4 |x|^4 }  \,\, \sum_C C^{AB}_C  \, \,\CO^C (0). 
\eeq
The coefficients $C^{AB}_C$ are a type of fusion rule.   
  For simplicity first consider only a single perturbation $\sum_A g_A \, \CO^A = g\, \CO$: 
\beq\label{ope.4} 
   \langle \X\rangle = \langle \X\rangle_0
   - 4\pi^2 g \int d^4x\,\langle \,\CO (x) \X \rangle_0 
   + \inv{2}\,(4\pi^2 g)^2 \int d^4x\int d^4y\,
   \langle \CO (x)\,\CO(y) \X \rangle_0 + \dots \,, 
\eeq
where the subscript ${}_0$ indicates that the correlation function 
is computed with respect to the free action $S_0$.   We have used that $\brabra \X \ketket = \langle \X \rangle$ since $\K |0\rangle = |0\rangle$.   
Using the OPE \eqref{ope.4} 
along with $\int_a d^4x/x^4=-2\pi^2\log a$, where $a$ 
is an ultraviolet cut-off, one finds
\beq\label{ope.5}
   \langle \X\rangle 
   = \langle \X\rangle_0 
   - 4\pi^2 (g+C g^2\log a) \int d^4x\,\langle \CO(x) \X \rangle
   + \dots \,. 
\eeq
The ultraviolet divergence is removed by letting 
$g\to g(a)=g-C g^2\log a$.   This leads to 
\beq
\label{betaA}
\beta_{g_A} \equiv \frac{dg_A}{d \ell}   = - \sum_{B,C} \, C^{BC}_A \, g_B g_C 
\eeq
where increasing $\ell = \log a $ is the flow to low energies.

Also of importance are  so-called ``scaling fields"  which are fields with a well-defined anomalous scaling dimension under RG flow, i.e. do not mix with other fields.    Throughout this paper,   $\gamma$ will refer to the anomalous correction to the naive classical  scaling dimension in mass units;
for instance for a mass parameter $m$ the scaling dimension is $1 + \gamma (m)$.   
Let $X(x)$ denote such a field having the following OPE with
the perturbation \eqref{perturbations} operators 
\beq\label{ope.9}
   \CO^A(x)\, X(y) = \frac{D^A}{8\pi^4 |x-y|^4}\, X(y) + \dots
\eeq
for some coefficients $D^A$.   Then to first order in $g$
\barray\label{ope.10}
   \langle X\rangle 
   &=& \langle X \rangle_0 
    - 4\pi^2 g_A \int d^4y\,\langle\CO^A(y)\, X \rangle_0  
    + \dots \nonumber\\
   &=& \( 1 + g_A D^A \log a \) \langle X\rangle_0 + \dots
    \approx a^{g_A D^A }\,\langle X \rangle_0 \,. 
\earray
 Thus $X$ has anomalous scaling dimension 
\beq
\label{gammaX}
\gamma (X)  =  \sum_A g_A D^A .
\eeq

To calculate the OPE coefficients $C^{AB}_C$ we need the OPE of the $J, \Jtilde$ operators.   $\K$-conjugation leads to an extra minus sign in the expansion of $\Phi^\daggerc$ in comparison with \eqref{as}:
\beq
\label{phidagc}
\Phi^\daggerc  (x)=  \int \frac{d^3 \kvec}{( 2 \pi)^{3/2}} \inv{\sqrt{2 \omega_\kvec}} \(-  a_-(\kvec) e^{i k\cdot x}  + a^\dagger_+ (\kvec) e^{-i k\cdot x } \).
\eeq
Using $a_\pm | 0 \rangle =0$  and $\K |0\rangle = |0\rangle$ one finds 
\beq
\label{onepoint}
\brabra  \Phi^\daggerc_i (x) \Phi_j (y) \ketket = - \brabra  \Phi_i (x) \Phi^\daggerc_j (y) \ketket  = \frac{\delta_{ij}}{4 \pi^2 |x-y|^2 },
\eeq
and similarly for $\brabra \Phitilde^\daggerc  (x) \Phitilde(y) \ketket$.   These two-point correlators,  i.e.  propagators,  satisfy the usual constraints of causality.  
Since the $\Phi$ and $\Phitilde$ fields are independent, 
\beq
\label{PhiPhitilde}
\brabra  \Phi^\daggerc_i (x) \Phitilde_j (y) \ketket =0,
\eeq
and this simplifies the computation of OPE coefficients $C^{AB}_C$.    
  Most importantly the  extra minus sign in \eqref{phidagc} leads to $[\tau^a , \tau^b]$ in the OPE of $J^a (x) J^b (y)$.\footnote{We refer to 
  \cite{RDolls2} for more technical details.}      
  In order for the RG fusion coefficients $C^{AB}_C$ to close,   we require that the $\tau^a$ comprise a Lie algebra: 
\beq
\label{taus}
[\tau^a, \tau^b ] = i f^{abc} \tau^c , ~~~~~ \tau^\dagger = \tau .
\eeq
Then one finds
\beq
\label{JOPE}
J^a (x) J^b (y) = - \frac{\kappa}{16 \pi^4 |x-y|^4 } {\rm Tr} ( \tau^a \tau^b ) -  \frac{i f^{abc}}{4 \pi^2 |x-y|^2 } \, J^c (y) + \ldots 
\eeq
In the above equation $\kappa=1$,  however we introduced $\kappa \neq 1$  for  book keeping reasons  to be explained below.   

The OPE \eqref{JOPE} has not been considered before and it is primarily for this reason that our model has no overlap with  more standard generalizations of $\Phi^4$ theory.   Since this OPE is central to the remainder of this article,  let make a few comments.    
Since $J^a (x)$ are quantum operators,  they do not necessarily commute.  If one simple exchanges $a,b$ and simulaneously $x,y$ on the LHS of \eqref{JOPE},  then it becomes 
$J^b(y) J^a (x)$  which is not equal to $J^a(x) J^b (y)$ according to the RHS of equation \eqref{JOPE} because of the anti-symmetry of the structure constant $f^{abc}$:
\beq
\label{notcommute}
  \lim_{x \to y} \, J^a (x) J^b (y)  \neq  \lim_{y \to x} \, J^b (y) J^a (x) .
  \eeq
 The above implies that the operators $J^a (x)$ are non-local,   which,  as explained above,    is a consequence of $\Phi^\daggerk$ being non-local.   
What is meaningful is that the operator equation \eqref{JOPE} is consistent with $\K$-conjugation,   and from this one can see why the  $i= \sqrt{-1}$ is necessary.    Taking the
$\K$-conjugate of both sides of \eqref{JOPE} one finds
\beq
\label{JOPEconj}
\( J^a (x) J^b (y) \)^\daggerk = J^b(y) J^a (x) = - \frac{2 \kappa \delta^{ab} }{16 \pi^4 |x-y|^4 }  +  \frac{i f^{abc}}{4 \pi^2 |x-y|^2 } \, J^c (y) ,
\eeq
and this is identical to  \eqref{JOPE} since $f^{abc} = - f^{bac}$.   

For euclidean QFT,      the perturbing operators of a CFT are normally  required to be local for physical reasons. 
   Below,   the marginal perturbations will be  
$\CO^a (x) =   J^a (x) \Jtilde^a (x)$  which are indeed local as is evident from the OPE in equation \eqref{OOPE}.
To help in  clarifying  this,    it is useful to make an analogy with free massless scalar fields in 2 spacetime dimensions,  especially in light of the comparison 
with 2D current-current perturbations.        In 2D,  the conserved current operators are in fact local,   however here the operators $J^a$ are not local,  nor are they  conserved currents.   
Consider a free massless scalar field in 2D with the local propagator $\langle \phi (x) \phi (y) \rangle = - \log |x-y|^2$.   In 2D CFT,  the field $\phi (x)$ is 
separated into left and right moving parts:   $\phi(x) = \varphi (z) + \bar{\varphi} (\bar{z})$  where $z = x + i y, ~ \bar{z} = x- iy$.   The two-point function 
is $\langle \varphi (z) \varphi (w) \rangle = - \log (z-w)$ and similarly for $\bar{\varphi}$. 
The field  $\phi$  is local,   but $\varphi$ and $\bar{\varphi}$ are non-local.    Consider a perturbation by  local operators $e^{i \alpha \phi} = e^{i \alpha \varphi} e^{i \alpha 
\bar{\varphi}} $.   Then the OPE of the holomorphic factor is non-local.   Using the Wick expansion,    
\beq
\label{scalarope}
\lim_{z \to w} \,  e^{i \alpha \varphi (z) } e^{i  \alpha \varphi (w) }  \sim (z-w)^{\alpha^2}  \, e^{2 i \alpha \varphi (w) } .
\eeq
The above OPE is obviously non-local in that it is not invariant under the exchange of $z$ and $w$,  analogous to the OPE \eqref{JOPE}.  
On the other hand the OPE  of  the local field $e^{i \alpha \phi (x)}$ with  $e^{i \alpha \phi (y)}$  is proportional to $|x-y|^{2 \alpha^2}$ which is local.

The perturbed theory is renormalizable,  at least to one loop,    if the operator algebra \eqref{OOPE} closes.    This places algebraic constraints on the allowed operators $\CO^A$ based on the structure constants $f^{abc}$.      Using \eqref{JOPE},   the theory is only 
renormalizable if 
\beq
\label{Closure}
d^A_{ab}\, d^B_{cd}\, f^{aci}\, f^{bdj} =-4  \sum_C C^{AB}_{C} \,d^C_{ij} .
\eeq
In practice,   if one starts with a single operator $\CO^A$,   then the above equation determines which additional operators $\CO^B$ with $B\neq A$ need to be included in the action in order for it to be renormalizable to 1-loop.    
As we show below,  based on the OPE  \eqref{JOPE},   our model is indeed  1-loop renormalizable. 
In particular there are no additional operators generated such as $J^a J^b$ nor $\Jtilde^a \Jtilde^b$.\footnote{{\bf Note added:}  It was argued that this is also the case to higher orders in \cite{RDolls2}.}  
In order to define scaling fields below,   we will also need the OPE's
\beq
\label{JPhiOPE}
J^a (x) \, \Phi_i (0) \sim \inv{4 \pi^2 |x|^2}  \sum_j \tau^a_{ij} \Phi_j (0), ~~~~~
J^a (x) \, \Phi^\daggerc_i (0) \sim  - \inv{4 \pi^2 |x|^2}  \sum_j \tau^a_{ji} \Phi^\daggerc_j (0) ,
\eeq
and similarly for $\tilde{J}, \Phitilde$.

The algebraic structure of the OPE of our model closely parallels current-current perturbations in 2D \cite{GLM}.     For instance the condition  \eqref{Closure} appears there.   
For the latter the OPE structure allowed us to propose an all-orders beta function.     In particular,  the $\kappa$-term in \eqref{JOPE} contributes at higher orders.
Whether this analysis can generalize to the models considered here  is an interesting question which is left for
 future study.\footnote{{\bf Note added:}   In our follow-up work \cite{RDolls2},    we carried this out to 3-loops and remarkably found that the beta functions for the model in this paper are  identical to those  for some analogous current-current perturbations in 2D. This allowed us to conjecture an all-orders beta function.}

\subsection{SU(2)  and its breaking to U(1)}

\subsubsection{Fully anisotropic case}

We now return to the case where the fields $\Phi_{i}, \, \Phitilde_{i}$,  $ i=1,2$ transform in  the fundamental representation of SU(2). 
We chose the $2\times 2$ matrices $\tau^a$ to be the hermitian Pauli matrices:
\beq
\label{tauasigma} 
\tau^a = \sigma^a, ~~~~~ \sigma^1 = \paulix,  ~~~\sigma^2 =  \pauliy,  ~~~\sigma^3 = \pauliz, ~~~{\rm Tr} ( \tau^a \tau^b) = 2 \delta^{ab} .
\eeq
For potential applications to Higgs physics,   we are mainly interested in the case where the SU(2) is broken to U(1) as in the Higgs mechanism,    however it is simpler to first find the beta functions in the fully anisotropic case,   and afterwards identify various couplings.   
For  fully anisotropic perturbations  in \eqref{perturbations}  we chose  $d^A_{ab} = d_{ab} = \delta_{ab}$: 
\beq
\label{beta1} 
\sum_A  \, g_A \CO^A = \half  \( g_1 \, \CO^1 +g_2 \, \CO^2 +g_3 \, \CO^3 \),   ~~~~ ~~~\CO^a \equiv J^a \Jtilde^a ,
\eeq  
with $g_1 \neq g_2 \neq g_3$.   
Using the OPE's \eqref{JOPE} and similarly for $\Jtilde^a$, one  finds
\beq
\label{Cs}
C^{12}_3 = C^{21}_{3} = C^{31}_{2} = C^{13}_2 = C^{23}_{1} = C^{32}_{1} = -1/2.
\eeq
This gives 
\beq
\label{FullyAniso}
\beta_{g_1} = g_2 g_3, ~~~~
\beta_{g_2} = g_1 g_3 ,~~~~
\beta_{g_3} = g_1 g_2.
\eeq

For these fully anisotropic perturbations,   the RG flows are simplified by certain RG invariants $Q$:
\beq
\label{Invariant}
\sum_g \beta_g \frac{\d Q}{\d g} = 0.
\eeq
There are $3$ such RG invariants:
\beq
\label{Invariant2}
Q_1 = g_2^2 - g_3^2, ~~~~Q_2 = g_3^2 - g_1^2 , ~~~~~Q_3 = g_1^2 - g_2^2 .
\eeq
However they are not all independent:   
\beq
\label{sumQs}
Q_1 + Q_2 + Q_3 = 0.
\eeq
Using two RG invariants one can reduce the RG flow to that of a single coupling,   for example $g_3$,   where 
$\beta_{g_3}$ is a function of $g_3$ and the RG invariants $Q_1, Q_2$.    Integrating the flow,   $g_3 (\ell)$ can be expressed in terms of 
Jacobi elliptic functions \cite{Elliptic},\footnote{See equations (57),(58) in \cite{Elliptic}.}    which are generally doubly periodic and exhibit {\it two}  distinct RG periods,   which suggests that the couplings have the topology of a torus.
For integrable theories in 2D with the same 1-loop beta functions 
\eqref{FullyAniso},    it was proposed in \cite{Elliptic} that the S-matrix is the one of Zamolodchikov \cite{Zelip},  which  transforms covariantly under $\SL2Z$.  (See Section V below.)


\subsubsection{Fully isotropic case}

For the fully isotropic case, $g_1 = g_2 = g_3 \equiv g$,   the perturbation is
\beq
\label{fullJJ}
\sum_{a=1}^3  g_a \CO^a  = \frac{g}{2}  \, \sum_{a=1}^3 J^a \Jtilde^a .
\eeq
The $J^a,  \Jtilde^a$ transform as the adjoint of SU(2) and the above field is built on the quadratic Casimir and thus SU(2)  invariant. 
The above beta function \eqref{FullyAniso} leads to   
\beq
\label{betaiso}
\beta_g = g^2 .
\eeq
      Thus for $g>0$ the perturbation is marginally relevant,  whereas   if $g<0$,  the perturbation is marginally irrelevant.  

Using the identity 
\beq
\label{sigmaIdentity}
\sum_a  \sigma^a_{ij} \sigma^a_{k\ell} = 2 \, \delta_{i\ell} \delta_{jk} -\delta_{ij} \delta_{k\ell} ,
\eeq
the perturbation can be written as 
\beq
\label{fullysymmetric}
\sum_a J^a \Jtilde^a =   2 (\Phi^\daggerc \Phitilde) (\Phitilde^\daggerc \Phi ) - (\Phi^\daggerc \Phi)(\Phitilde^\daggerc \Phitilde ) 
\eeq
Note that for $\Phitilde = \Phi$ this is the standard Higgs potential apart from $\dagger$ replaced with $\daggerc$.

Let us consider possible quadratic mass terms.     The composite fields  $\Phi^\daggerc \Phi$ and $\Phitilde ^\daggerc \Phitilde$ are {\it not}  scaling fields, 
whereas the mixed operators  $\Phi^\daggerc \Phitilde$ and $\Phitilde^\daggerc \Phi$ are.     This follows from the OPE's
\beq
\label{scaling1}
\CO^a (x) \, [\Phi^\daggerc \Phitilde (0)]  \sim   - \inv{16 \pi^4 |x|^4} [\Phi^\daggerc \Phitilde (0) ] , ~~~~~
\CO^a (x) \, [\Phitilde^\daggerc \Phi(0)] \sim   - \inv{16 \pi^4 |x|^4} [\Phitilde^\daggerc \Phi (0)] , ~~~ \forall a,
\eeq
 where we have used \eqref{JPhiOPE} and $(\sigma^a)^2 = 1$ $\forall a$.    Using \eqref{gammaX} one finds to one loop 
 \beq
 \label{gammaPhiPhitilde}
 \gamma (\Phi^\daggerc \Phitilde ) =    \gamma (\Phitilde^\daggerc \Phi) = - \frac{3g}{4}.
 \eeq
We thus consider the possible mass terms: 
\beq
\label{Smass}
\CS_{\rm mass-terms} = \int d^4 x  \,\,  \frac{m^2}{2} \( \Phi^\daggerc \Phitilde + \Phitilde^\daggerc \Phi \).
\eeq
Using the fact that the action must be dimensionless implies
$\[2 + \gamma(m^2) \]+ \[2 + \gamma(\Phi^\daggerc \Phitilde)\] = 4$.
Thus 
 \beq
 \label{gammaMass}
 \gamma (m^2) = \inv{m^2} \frac{d m^2}{d \ell} =  \frac{3g}{4}. 
 \eeq


\subsubsection{SU(2) broken to U(1)}

\begin{figure}[t]
\centering\includegraphics[width=.5\textwidth]{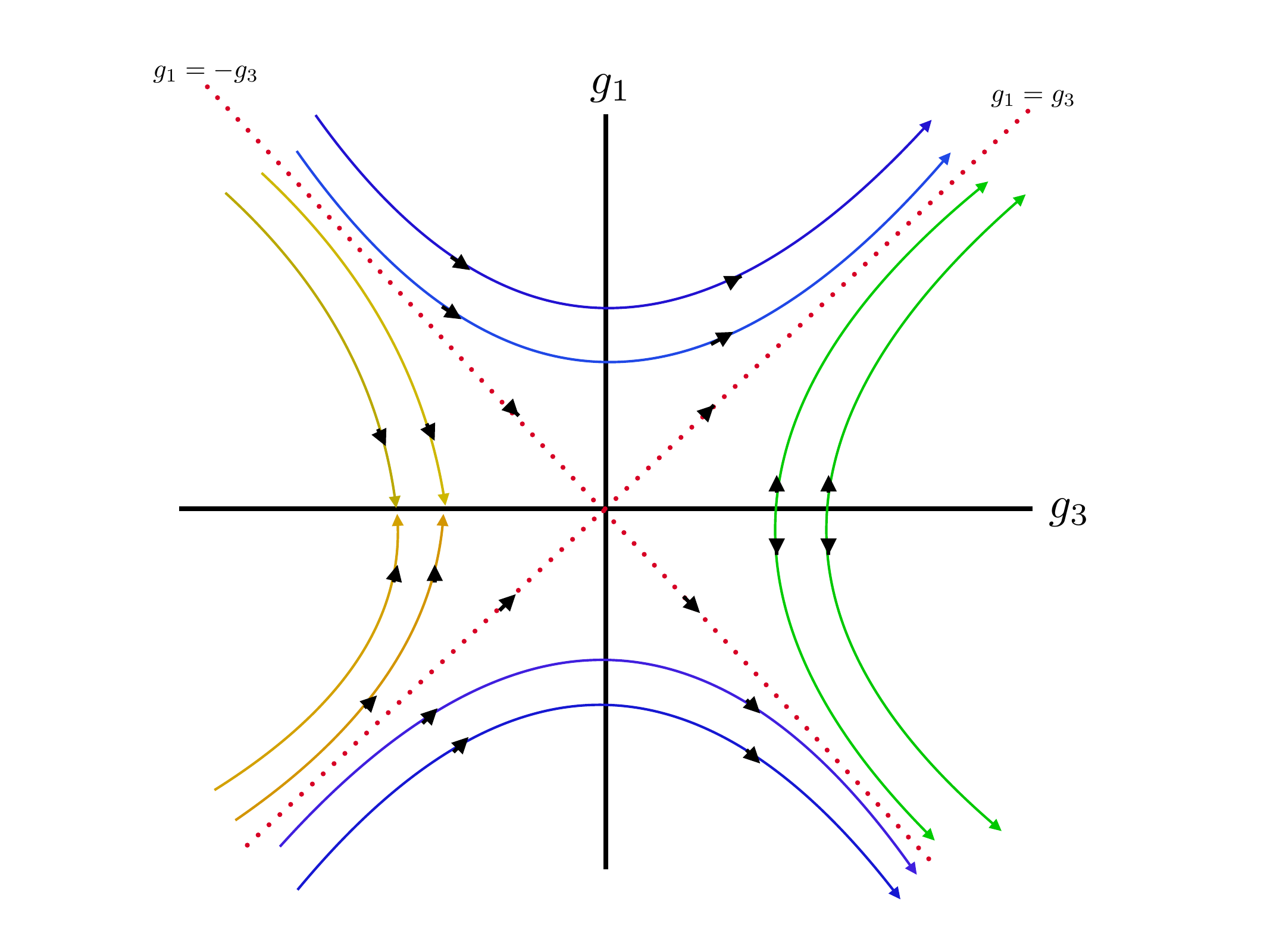}
\caption{RG flows based on the beta-function \eqref{NotFully1}.   RG trajectories are hyperbolas $g_1^2 - g_3^2 = Q = {\rm constant}$.  Arrows refer to the flow to low energies.    $g_1 = 0$  represents a line of RG  fixed points,  where green curves are relevant perturbations which originate from UV fixed points,  and yellow curves represent IR fixed points.  
 The blue curves are cyclic RG flows.}
 \label{Flows}
\end{figure}

The above fully anisotropic case is rather complicated even for  2D  current-current perturbations  (see further remarks in Section V where we review how  the fully anisotropic case in 2D  leads to S-matrices expressed in terms of Jacobi elliptic functions).
A simpler and potentially more physically  interesting case is to break the SU(2) symmetry to U(1) $\subset$ SU(2).
Setting $g_1 = g_2$, 
\beq
\label{NotFully1}
\beta_{g_1} = g_1 g_3 , ~~~~\beta_{g_3} = g_1^2 .
\eeq
There is now only a single RG invariant: 
\beq
\label{Q}
Q = g_1^2 - g_3^2. 
\eeq
The RG flow trajectories are thus hyperbolas $Q = {\rm constant}$,    as shown in Figure \ref{Flows},  where 
 $g_1 = 0$ represents a line of fixed points.\footnote{{\bf Note added:}   Taking $g_1$ to be purely imaginary,   the 
 RG flows become  circles that connect two different points along the $g_3$ line of fixed points,   which correspond to massless flows between UV and IR fixed points \cite{RDolls2}.}   Along the SU(2) symmetric separatrices  $g_1 = \pm g_3$,  one has a single fixed point $g_1 = g_3$ in the UV or IR.    Just below the separatrix,   flows either begin or terminate along the line of fixed points $g_1 =0$,    where 
 $g_3>0$ is marginally relevant,  whereas $g_3<0 $ is marginally irrelevant.  
 On the other hand,  just above the separatrix the flows have no fixed points and are in fact cyclic,  as we will show below.
 
 One can easily find the anomalous dimension of the perturbation along the line of fixed  points.    On general grounds,  given a
 beta function for one coupling $g$ with fixed point $\beta(g_c) =0$.    Expanding around this fixed point in $D=4$ spacetime dimensions:
 \beq
 \label{betacritical} 
 \beta_g \sim  (4 -\Gamma_{\rm pert} ) (g-g_c),
 \eeq
where $\Gamma_{\rm pert}$ is the scaling dimension of the perturbation.      For the coupling $g_1$, one has   $g_{1c} =0$ and from the slope of the 
beta function $\beta_{g_1} = g_1 g_3$ to one loop,  one finds 
 \beq
 \label{Gammpert}
 \Gamma_{\rm pert} = 4 -g_3,    ~~  \Longrightarrow ~~  \gamma (\CO^1) =  \gamma(\CO^2) =  - g_3. 
 \eeq

If $Q<0$,   i.e. below the separatrices $g_3 = \pm g_1$,    then the RG flows either originate or terminate along the line of fixed points. 
On the other hand if $Q>0$  the flows have no fixed points anywhere  and are in fact cyclic.        This is easily seen as follows.
The coupling $g_1$ can be eliminated and one has 
\beq
\label{NotFully2} 
\frac{d g_3}{d \ell} = g_3^2 + Q,
\eeq
which is easily integrated:
\beq
\label{g3ofell}
g_3 (\ell) = \sqrt{Q} \, \tan \( \sqrt{Q} ( \ell - \hat{\ell}_0 ) \), ~~~ g_1 (\ell) =  \frac{\sqrt{Q}}{\cos \( \sqrt{Q} ( \ell - \hat{\ell}_0 ) \)}
\eeq
where $\hat{\ell}_0$ is an integration constant.
From this one  identifies the RG  period $\lambda$ which is RG invariant since $Q$ is:
\beq
\label{period}
g_3 (\ell + \lambda ) = g_3 (\ell), ~~~~ \lambda = \frac{\pi}{\sqrt{Q}} .
\eeq
In contrast,    for $Q<0$,   the flows \eqref{g3ofell} reach,  in the UV or IR,  the line of fixed points $g_1 =0$.

The above analysis is based only on the 1-loop beta functions and one should question whether the above cyclic RG flows 
are spoiled at higher loops.\footnote{{\bf Note added:}  This was carried out to 3-loops in \cite{RDolls2},  where it was shown that the 1-loop features of the RG flows  persist.}
      For the case of current-current perturbations in 2D  it was shown in \cite{BLflow} that these cyclic flows persist to all orders in the coupling.   There continues to be an RG invariant $Q$ and the RG period $\lambda$ is still a function of $Q$.    The higher order beta functions merely double the RG period,  so that 
\eqref{period} becomes $\lambda = 2 \pi/\sqrt{Q}$.    The higher loop corrections to $Q$ can be found in \cite{BLflow},  and are repeated
below in  equation \eqref{Qallorders}.      It was argued that the space of couplings $g_1, g_3$ has the topology of a cylinder,   where $g_3 = \pm \infty$ are identified.
For the fully anisotropic case \eqref{FullyAniso},   since the integrated beta functions have two periods based on the Jacobi elliptic functions,   we suggest  that  in this case the topology of the space of couplings  is that of a torus.   
This suggests that the RG flows can be uniformized using elliptic functions, reminiscent of Seiberg-Witten theory \cite{SeibergWitten}.   
Let us add that the integrated RG flow for the beta functions \eqref{NotFully1} involves trigonometric functions,    as does the S-matrix for the cyclic sine-Gordon model reviewed in the next section,  whereas  for the fully anisotropic case,   the integrated couplings and S-matrix involve elliptic functions.     
Furthermore,   the all-orders beta functions in \cite{GLM,BLflow} have a $g \to 1/g$ duality,  suggesting that the coupling constants may have an
${\rm SL} (2, \Z)$  modular symmetry,  or a subgroup of it.       We will comment on this further below,   but this is largely a  topic left for further study.

\section{Nested Russian Dolls as a signature of a cyclic RG flow in  the 2D case}

Above we have defined some novel marginal operators involving two Higgs doublets which led to a potentially rich structure of RG flows,  including a line of fixed points and a regime with cyclic RG flows.   
 For the remainder of this article,   we focus on the cyclic regime.\footnote{{\bf Note added:}   The CFT's along the line of fixed points are  also interesting in their own right and developed further in \cite{RDolls2}.     It was shown there that by taking 
 $g_1$ to be purely imaginary,   there exists flows connecting two non-trivial fixed points in the UV and IR.}     
To fully understand the physical manifestations of a cyclic RG will be a challenge,   especially in connection with spontaneous symmetry breaking and the Higgs mechanism.   We will return to this in Section VI.     
In order to gain some insight,  it is useful to consider some relativistic models in 2 spacetime dimensions which have 
{\it identical} beta functions.\footnote{{\bf Note added:}   In \cite{RDolls2}  it was argued that the higher order beta functions are the same as for the current-current perturbations in \cite{GLM}.}
 In this section,  we  briefly review how the cyclic RG is manifested in the spectrum and S-matrices for integrable models in 2D with same beta functions.     As explained in the Introduction,    a generic feature is an infinite spectrum of resonances,  in some cases unstable,    with the Russian Doll scaling behavior \eqref{ERussianDoll}.
 In general,  these models have an infinite Russian Doll spectrum,   which can be stable or unstable depending on the region of parameters.    
 For a general discussion of unstable particles for integrable QFT in 2D see \cite{Mussardo,mira2,Fring}.

\subsection{General properties:   Infinite spectrum of resonances with Russian Doll scaling behavior}

In 2D  it is conventional to express the energy and momentum of a single particle in terms of the rapidity $\theta$,   in part since the analytic properties of the S-matrix are more transparent in this variable:
\beq
\label{rapidities}
E = m \cosh \theta , ~~~~~p = m \sinh \theta .
\eeq
Since there is no particle production,   the S-matrix is only a function of the Mandelstam variable $s$:
\beq
\label{stheta}
s = (k_1 + k_2)^2 = 2 m^2 \, \cosh^2 (\theta_{12}/2), ~~~~~\theta_{12} \equiv  \theta_1 - \theta_2 .
\eeq

For the cyclic sine-Gordon model reviewed below,     real analyticity  is violated  $S^\dagger (\theta_{12} ) \neq S(-\theta_{12})$.   
However it still defines a unitary time evolution if one replaces real analyticity with Hermitian analyticity \cite{Olive,Eden,Miramontes1}.
There are general arguments that this can  occur  when the theory breaks time-reversal $\CT$ or parity $\CP$.   
To illustrate,   consider diagonal scattering with particles of type $a,b$.      If parity is broken,   then 
\beq
\label{HA2}
S_{ab} (\theta_{12} ) \neq S_{ba} (\theta_{12} ), ~~~~~S_{ba} (\theta_{12} ) =  S^*_{ab} (\theta_{21} ). 
\eeq
Unitary then corresponds to 
\beq
\label{Unitarity}
S^\dagger S = 1 ~~~~~~~\Longrightarrow ~~~~~S^*_{ba} (\theta ) S_{ab} (\theta) = S_{ab} (\theta) S_{ab} (-\theta) = 1. 
\eeq
This suggests these features should  also  be a property of the pseudo-Hermitian theories in the previous sections,  and this is indeed the case for the cyclic sine-Gordon model.

The S-matrices for  QFT's with beta-functions  \eqref{NotFully1} and \eqref{FullyAniso} were proposed in \cite{LeClairSierra} 
and \cite{Elliptic} respectively,   and will be reviewed below.       In both cases,   the cyclic RG is manifested by an infinite number of resonances 
corresponding to the poles in $s$.    For simplicity consider one such pole at 
\beq
\label{poles}
s = (m - i \Gamma/2 )^2 .
\eeq
If $\Gamma \ll m$,   The S-matrix has the Breit-Wigner form near the resonance:
\beq
\label{poles2}
S \simeq 1 - i \frac{2 m \Gamma}{s-m^2 + i m \Gamma}.   
\eeq
These poles in $s$ correspond to poles at $\theta = \mu - i \eta$,   which lead to 
\barray
m^2 - \Gamma^2/4 &=& 2 m^2 \( 1 + \cosh \mu \cos \eta \) 
\\
m \Gamma &=& 2 m^2 \, \sinh \mu \sin \eta .
\label{Breit2}
\earray

\def\msg{m}

\def\Msoliton{M_s}

\subsection{Trigonometric case:   Cyclic sine-Gordon model}

This model is based on current-current perturbations with the beta functions \eqref{NotFully1},   which can be mapped to an analytic continuation of 
the sine-Gordon model. 
Let us first briefly review the usual sine-Gordon model.     It  can be defined by the action 
\beq
\label{sGaction}
\CS = \inv{4 \pi} \int d^2 x \( \half (\d \phi)^2 + \frac{\msg^2}{b^2}  \,  \cos b \phi \).
\eeq
Above,  $\msg$ is the classical mass of the field $\phi$.     The $\cos b \phi$ term has anomalous dimension $b^2$ which is relevant for 
$b^2 < 2$.   For real $1< b^2 < 2$,  the model only has solitons of mass $\Msoliton$ which carry a topological $U(1)$ charge.  The S-matrix for these solitons is non-diagonal  and  was  determined in \cite{ZZ}  for arbitrary $b$,    where the field $\phi$ corresponds to a bound state of 2 solitons when $b^2 < 1$,  the so-called first ``breather".   When $b^2 = 1$,   the theory is a free Dirac fermion,   where  under bosonization the 
cosine  term is just a mass term.    Based on the scaling dimension of $\cos b \phi$,  in the quantum theory $\msg$ has anomalous dimension 
$(2-b^2)/2$,   thus the physical mass $\Msoliton \propto \msg^{2/{(2-b^2)}}$.       

In \cite{LeClairSierra} the cyclic regime of current perturbations found in \cite{BLflow},   which to one loop has the same beta function as in 
\eqref{NotFully1},    was mapped into the sine-Gordon model at the coupling 
\beq
\label{bcyclic}
b^2 =  \frac{2}{1 + i \pi /\lambda},
\eeq
where $\lambda$ is the RG period.   The above  \eqref{bcyclic}  is a complex deformation of the marginal point $b^2 =2$.  
This map is mainly based on the existence of an RG invariant  $Q$ to all orders in the couplings $g_1, g_3$ which can be found in 
\cite{BLflow}.\footnote{It was found that the relation between the RG period 
$\lambda$ and the non-perturbative $Q$  differs by a factor of $2$ in comparison with
\eqref{period},  namely $\lambda = 2 \pi \sqrt{Q}$,  however this is unimportant for the main considerations of this article.}       
The usual sine-Gordon theory is known to have  affine quantum group symmetry $\CU_q  \hat{SU(2)}$ with $q = e^{2 \pi i /b^2}$ \cite{BLnonlocal}.      Whereas $q$ is a phase for the usual sine-Gordon model,  for $b$ in \eqref{bcyclic} the quantum group deformation parameter  $q$ is real, 
$q = - e^{- 2 \pi \sqrt{Q} } = - e^{- \pi^2/\lambda}$.        The completion of the Figure \ref{Flows} to strong coupling can be found in \cite{BLflow}.   

The S-matrix for the  cyclic sine-Gordon model is defined by the analytic continuation of the usual sine-Gordon soliton S-matrix to the complex coupling in \eqref{bcyclic}, as proposed in \cite{LeClairSierra},  and the S-matrix is expressed in terms of trigonometric functions.      Referring to \cite{LeClairSierra} for details,     in was found there that this S-matrix 
has an infinite Russian Doll spectrum of resonances,   a clear signature of a cyclic RG.        These resonances correspond to the following poles in rapidity:  
\beq
\label{cyclicsG1}
\theta_n = n \lambda, ~~~~n = 1, 2, 3, \ldots , ~~~~~~~~\Longrightarrow ~~ m_n = 2 \Msoliton  \cosh (n \lambda/2 ). 
\eeq
There are also poles in the crossed channel corresponding to an imaginary mass: 
\beq
\label{cyclicsG2}
\bar{\theta}_n = i \pi - n \lambda, ~~~~n = 1, 2, 3, \ldots , ~~~\Longrightarrow ~~ m_n = 2 i  \Msoliton   \sinh  (n \lambda/2 ). 
\eeq
Note that for this model,   $\eta = 0$ so that the  resonances in \eqref{cyclicsG1} are stable,  i.e. $\Gamma = 0$.



\subsection{Elliptic examples and $\SL2Z$} 

The first S-matrix built out of elliptic functions was found by Zamolodchikov \cite{Zelip}.   In that work an underlying QFT was not proposed,    however it was suggested  that the periodicity as a function of energy in some regimes could perhaps be interpreted as a consequence of a  cyclic RG.    An underlying QFT for this S-matrix was finally proposed in 
\cite{Elliptic} based on the fully anisotropic SU(2) current-current perturbations  with the fully anisotropic beta function \eqref{FullyAniso}.     Since the elliptic functions have two parameters,   it has a number of interesting limits,  including the cyclic sine-Gordon model reviewed above.      There is also a region 
of parameters such that the appropriate  limit is the diagonal,  single particle S-matrix built on elliptic functions studied by Mussardo and Penati \cite{Mussardo},  which in principle provides an underlying QFT description of the latter.    

Zamolodchikov's S-matrix in \cite{Zelip}  is primarily characterized by a  $\Z_4$ symmetry.     $\Z_4$  is a subgroup of $\SL2Z$ generated by the matrix  $A = \begin{pmatrix} 0 & -1 \\ 1 & 0 \end{pmatrix} $.
This elliptic solution of the Yang-Baxter equation is  defined on an elliptic curve with modulus 
$\tau$,   and though the solution to the Yang-Baxter equation is not strictly  $\SL2Z$ invariant based on the transformations of the elliptic functions involved,  they are  covariant under $\SL2Z$,  meaning the S-matrix transforms with phase factors or gauge transformations that preserve the structure of the Yang-Baxter equation \cite{Baxter,Belavin}.

\def\PhiVev{\langle \Phi \rangle}
\def\PhitildeVev{\langle \Phitilde \rangle}


\section{Spontaneous Symmetry Breaking:   Infinite number of VEV's with Russian Doll scaling}

 A detailed study of  the interplay between the Higgs mechanism and a cyclic RG  is clearly interesting theoretically,   
 however is beyond the original scope of this article,   since the original motivation was to simply find interesting RG flows for some new Higgs-like potentials with a Lie-algebraic structure. 
 Nevertheless,   since our model with $\Phi = \Phitilde$ is the same as for a standard Higgs potential,   except for 
 $\dagger$ replaced by $\daggerk$,    it would be remiss not to  consider spontaneous symmetry breaking (SSB) for our model.  
 Above we have only described an {\it explicit breaking}  of SU(2) to U(1) rather than SSB,   and the resulting beta functions.  In this section we present arguments that this resulting beta function is  meaningful {\it after}  SSB.

 We start with the fully isotropic case $g_1 = g_2 = g_3 = g$ considered above.
 SSB  requires adding mass terms that are consistent with the symmetries.
We also require them to be scaling fields as in \eqref{Smass}.     Let us first assume that these terms are negative,   however as we will see,  
due to the cyclicity of the RG,  one does not have to assume this since the coupling $g$ will change signs in each RG cycle.  
We thus consider the Higgs-like potential:
\beq
\label{Vpotential}
V(\Phi, \Phitilde) = - \frac{m^2}{2} \( \Phi^\daggerc \Phitilde + \Phitilde^\daggerc \Phi \) + \frac{g}{2} \sum_{a=1}^3  J^a \Jtilde^a .
\eeq
Minimizing the potential with respect to the fields one obtains two equations 
\barray
- m^2 \PhitildeVev + g \sum_a \sigma^a \PhiVev \, \langle \Jtilde^a \rangle &=0& \\
- m^2 \PhiVev + g \sum_a   \sigma^a \PhitildeVev \, \langle  J^a \rangle &=0,&
\earray
and their $\K$ conjugates,    
where $\langle \Phi \rangle \equiv \brabra \Phi \ketket $ denotes the vacuum expectation value (vev).   
There exists solutions with $\PhiVev = \PhitildeVev$ due to the $\Z_2$ symmetry that exchanges
$\Phi$ and $\Phitilde$.   Inserting the identity $\sum_a  \sigma^a \sigma^a = 3$ into the $m^2$ term above,   one sees that SSB requires 
a solution 
\beq
\label{SSB1} 
\sigma^a \PhiVev  \neq 0, ~~~~~ {\rm for~ some~ }a \in \{1,2,3\}.
\eeq
 Without loss of generality let us chose $\sigma^a = \sigma^3$ in the above equation and specify the vev 
 as 
 \beq 
 \label{SSB2}
 \sigma^3 \PhiVev =  \begin{pmatrix} 
 v \\ 0 
 \end{pmatrix},
 \eeq
 where $\Phi = \begin{pmatrix} 
 \Phi_1 \\ \Phi_2
 \end{pmatrix}$, 
 and we require $v$ to be real.   
 Then 
 \beq
 \label{SSB3}
 \langle J^3 \rangle = \langle \Phi^\daggerc_1  \Phi_1 \rangle = v^2 = \frac{m^2}{3g}.
 \eeq
 Since $v^2$ is positive,   there aren't solutions $v^2$  for every $m$ and $g$ in the above equation,  and this will be important below.

  \def\lvev{\ell_{{\rm vev};n}}
 
 \def\mfamily{\mathfrak{m}}

 SSB solutions for $\Phi = \Phitilde$ closely parallel  those  for the standard Higgs sector,     though  perhaps in an algebraically more transparent manner,   but with $\dagger$ replaced by $\daggerc$.       What is definitely different is  the cyclic behavior of the coupling $g$ after SSB.  
Consider expanding about the above minimum of the potential $J^3 = \langle J^3 \rangle$.     Then under renormalization,  the $J^1 (x) J^2 (0)$ OPE will reintroduce the operator $\CO^3$ with with a different coupling $g_3 \neq g_1 = g_2$.     Thus the cyclic flows in \eqref{g3ofell}  become relevant.   
 In the vev \eqref{SSB3} it is thus natural to take $g = g_1$ in that formula,  and since $g_3$ is perturbatively generated under RG,   it is natural to expect that $Q>0$ which leads to the cyclic regime.     
 One thereby  obtains 
 \beq
 \label{vev1}
 v^2 (\ell) = \frac{\lambda}{3 \pi} m^2 (\ell) \cos\( \frac{\pi}{\lambda} (\ell - \hat{\ell}_0)\).
 \eeq
 Since $v$ is real,   there are only solutions to the vev $v$ if the cosine  above is positive if $m^2$ is positive.   Had we started with 
 an initially negative $m^2$,   then here also there are solutions when the  cosine factor is negative. 
    Therefore there are SSB solutions $v$ whatever the initial sign of $m^2$ in the mass terms.      In this sense,  the theory spontaneously breaks  itself without having to justify a negative $m^2$ term from the beginning.      
 Let us assume $m^2$ is positive for concreteness.    Without loss of generality we set $\hat{\ell}_0 =0$,    and choose the physical region to be 
 $\ell> \lambda/2$.      Significant points are $\ell_m$ where $v=0$:
 \beq
 \label{vev2}
 v (\ell_m) = 0, ~~~~ \ell_m \equiv (m + \half) \lambda,    ~~~~ m=0,1, 2, \ldots
 \eeq
 There exists real positive solutions $v$ if $\ell_m < \ell < \ell_{m+1}$ with $m=1,3,5, \ldots$ odd.     For $m$ an even integer there are no solutions.  
 We define $\lvev$  as the $n$-th  solution to $v$ based on \eqref{vev1}.    There is a one parameter family of such solutions that depend on 
 an arbitrary angle $\vartheta_n$:
\beq
\label{vev3} 
\lvev = \ell_{2n-1} + \frac{\lambda}{\pi} \, \vartheta_n =  2n \lambda +  \lambda \( \vartheta_n /\pi - \half\),    ~~n=1,2,3,\ldots, ~~~     0< \vartheta_n <\pi/2 .
\eeq
For simplicity we take $m(\ell)$ to be determined by its classical scaling $m(\ell) = M e^\ell $.  
One then obtains the following for the $n$-th vev corresponding to the $n$-th RG cycle:  
 \beq
 \label{vev4}
 v_n^2 = v^2 ( \lvev ) =  \frac{\lambda \sin \vartheta_n}{3 \pi}  \,M^2 \,  \exp \( {4n\lambda + 2 \lambda (\vartheta_n - \pi/2  )/\pi }\) .
 \eeq
 
The physical significance of all these Russian Doll vev's $v_n$  is not entirely clear.   A physical  theory should have a single ground state,   not an infinity of them. 
Choosing the lowest one,   then based on the 2D models with the same beta function described above,   these additional vev's  could appear as higher unstable resonances in the S-matrix.       
 In the Russian Doll BCS model of superconductivity,  which breaks time-reversal symmetry,    there  are different BCS condensates  that arise from the cyclic RG and are better understood  \cite{LeClairSierraBCS}.  There,   the cyclic RG leads to states $E_n$ with the Russian Doll scaling 
as in \eqref{ERussianDoll}.

 \section{Speculation on the origin of families}

  In the  Standard Model of particle physics,   fermions acquire their masses through Yukawa couplings with the Higgs and these masses are proportional to the single standard Higgs vev.  
 In this section we explore the possibility that a cyclic RG may be at the origin of the existence of families in the Standard Model.   We do not attempt  to  prematurely build a complete model that is fully consistent with the experiments, 
 the main reason being  
  the many stringent experimental constraints confirming the Standard Model that need to be scrutinized,  and we lack the expertise to do this.   
  To repeat the idea stated in the Introduction,  
  we speculate  that each family is one of an infinite number of Russian Dolls with nearly identical properties apart from the large disparity in their masses.   
 
The first family of quarks and leptons are the SU(2) doublets 
\beq
\label{doublets} 
\updoublet  ,\edoublet .
\eeq
If this were the only family,    the theory is strongly constrained by the fundamental principles of gauge symmetry and the Higgs mechanism.    
What is mysterious is that the above family repeats itself with ever increasing masses leading to many more unexplained free parameters,  the masses themselves,    
and  the CKM matrix which incorporates mixing between families.    These are the two additional  families $\charmdoublet,\!\muondoublet$ and $\topdoublet,\!\taudoublet$.     There are no symmetry arguments for the existence of these additional families,   and this is part of their mystery.   
Needless to say,   the hierarchy and structure of Yukawa couplings are a major open question.

    The additional families have essentially identical  properties as far as gauge couplings,   and this suggests the existence of families has an origin that is separate from the gauge symmetry,   and such an explanation could come from the Higgs sector,  which is the least understood aspect of the Standard Model.   
 One way to characterize how the different families are nearly identical  copies is based on 't\,Hooft anomaly cancelation
 \cite{Anomalies,Bonora}.      The various quantum numbers of the first family are such that all gauge anomalies precisely cancel,  and this is a non-trivial matter.    More importantly,    the anomalies cancel for each individual family separately,   which is an indication that they are somehow copies of the first family,   apart from their masses.     The Higgs sector does not contribute to these anomalies,   which is also important for our considerations.   There is already a ``small hierarchy" issue,    in that the ratio of the top quark to up quark mass is about $10^5$ which is not understood;     it is just a free parameter in the Standard Model attributed to the strength of Yukawa couplings to the Higgs field.\footnote{It's even worse if one considers the ratio of say the top quark mass to that of neutrinos,  which is about $10^{12}$.}   
  Below we only focus on the increasing masses of the families to see if our  idea is at all plausible.

 In the Standard Model,   the fermions,  namely  the quarks and leptons,  obtain their masses from Yukawa couplings to the Higgs,  $ \sim y_f \, \Phi \,  \bar{\psi} \psi$,   and this leads to 
 fermion masses proportional to the vev  $v$,   $m_f \sim  y_f \, v$,  where $y_f$ is the  Yukawa coupling.     For the usual Higgs mechanism,   there is only one vev,   in contrast to the infinite number of vevs  $v_n$ found above.   The conventional view is that the mass hierarchy of families arises solely from the Yukawa couplings $y_f$ with the mass scale set by the single vev $v$.         In principle,   a theory has one ground state.    Let us identify this ground state with the lowest 
 vev,  $v_1$,  which gives masses to the first family \eqref{doublets}.   We then interpret the more massive families 
 as arising from the higher energy vev's,    as excited,  unstable particles  with higher mass proportional to $v_n$,   however we cannot yet prove this.   
 This is the weakest aspect  of our scenario,  which needs further justification,   but let us proceed regardless.   These are not excited states of the usual kind,  since here they arise from the cyclicity of the RG and all have the same quantum numbers.       From this point of view,  by analogy with the Russian Doll signature of resonances for 
 the 2D models with identical beta functions  reviewed in Section V,    the additional families  have identical gauge quantum numbers and could appear as resonances in scattering experiments. 
   For simplicity we do not consider in detail such Yukawa couplings,   but simply explore generic features that should be expected if the RG is indeed cyclic and the above reasoning holds.

 \subsection{Rough estimates of the RG period $\lambda$  and an argument for  the existence of only 3 families}

 Based on the above discussion,  let us  {\it roughly} identify the mass of the $n$-th family based on $v_n$ above.      
 Since we cannot predict the angles $\vartheta_n$,  let us take them all to be $\vartheta_n = \pi/2$, again for simplicity.     The mass $\mfamily_n$ of the 
 $n$-th family is then approximately 
 \beq
 \label{matterfit}
 \mfamily_n \approx  \sqrt{\frac{\lambda}{3 \pi}} \, M \, e^{2n \lambda} .
 \eeq
The fundamental parameter,  the RG period  $\lambda$, is determined by $\lambda = \pi/\sqrt{Q}$,   where  $Q= g_1^2 - g_3^2$ to 1-loop.  It cannot be predicted at this stage but can in principle be  inferred from experimental data if a cyclic RG is indeed operative.        Roughly speaking,   if the coupling $g_1$ is of order $1$ then $\lambda \sim  \pi$,   and as we will see below this is  roughly consistent with  data on masses of families.   A cleaner and more accurate estimate of the RG period $\lambda$ will be presented in the next subsection based on the phenomenological Koide formula.  
 
\begin{figure}[t]
\centering\includegraphics[width=.4\textwidth]{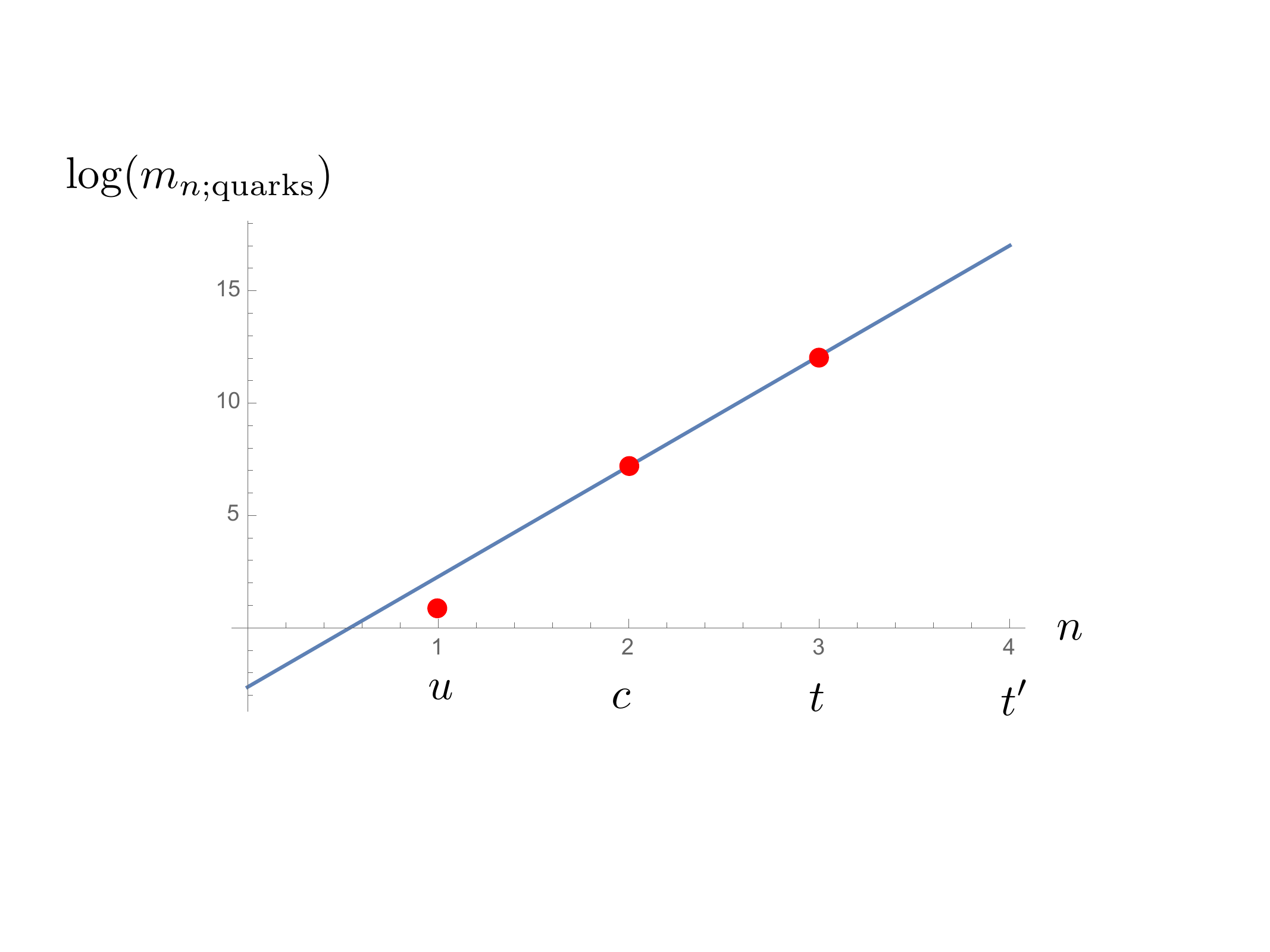}
\caption{Fit to quark  masses \eqref{QuarkFitExp}.  The red dots are the known physical masses.}
 \label{QuarkFit}
\end{figure}

\begin{figure}[t]
\centering\includegraphics[width=.4\textwidth]{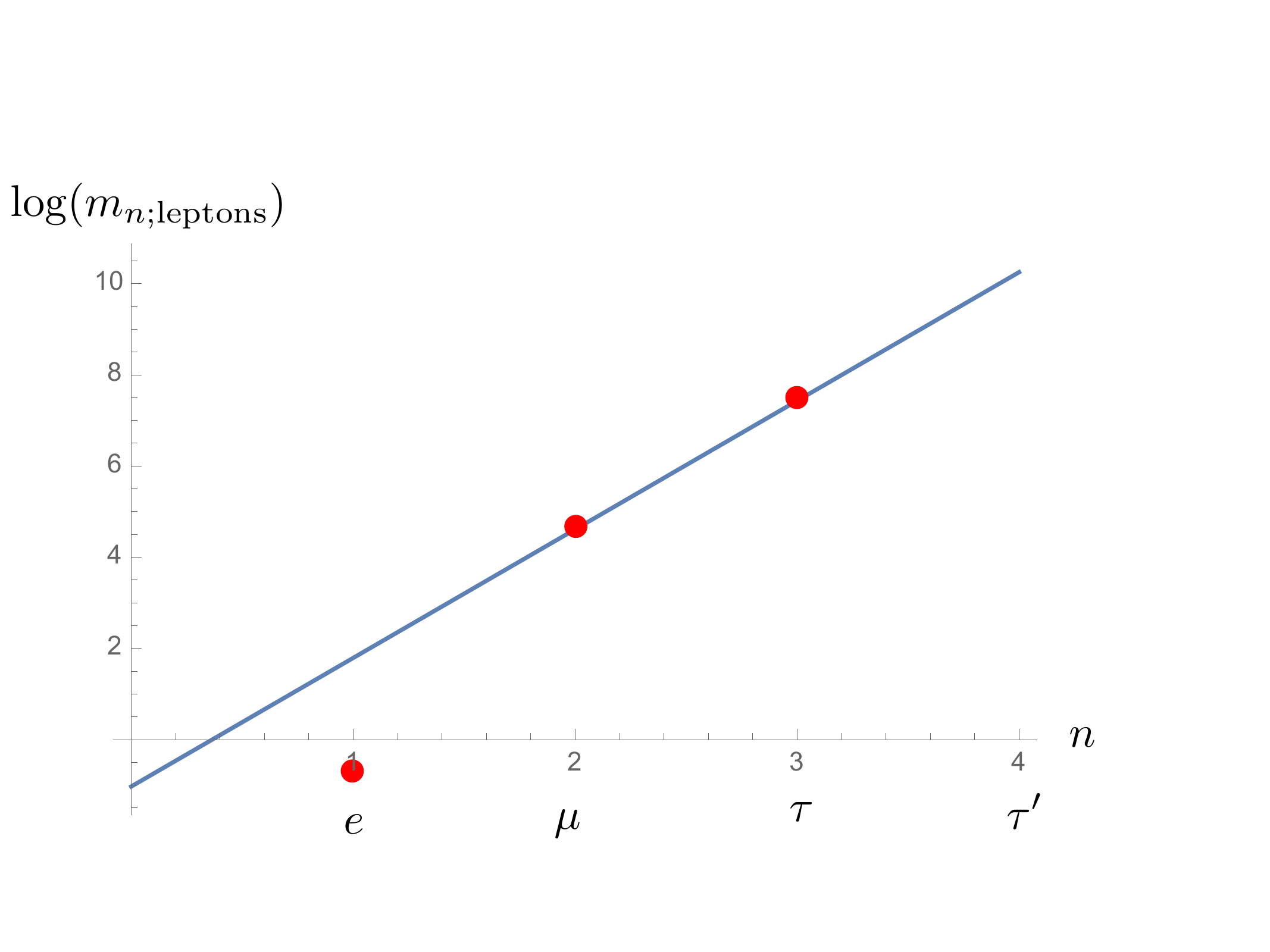}
\caption{Fit to lepton  masses \eqref{LeptonFitExp}.  The red dots are the known physical masses.}
 \label{LeptonFit}
\end{figure}

Let us make some crude checks of the above idea based on the known data on quark and lepton masses \cite{ParticleDataGroup}.  
We begin with the quarks since it is in this sector that the experimental bounds on a 4-th family are strongest.   
The masses of the upper component of the SU(2) doublets are  
$\{ m_u , m_c, m_t \}  \approx \{ 2.3, 1275. ,  173210. \} {\rm MeV }$. 
In spite of having only 3 data points,     fitting this data to \eqref{matterfit}  (with some embarrassment,  however we will do better in the next subsection)  we find 
\beq
\label{QuarkFitExp}
\mfamily_{n; {\rm (up)quarks}}  =  \sqrt{\frac{\lambda_{\rm qu}}{3 \pi} } \, M_{\rm qu} \,\,   e^{2 n \lambda_{\rm qu}},  ~~~~ M_{\rm qu} = 0.14 \, {\rm MeV}, ~~~~ \lambda _{\rm qu} = 2.4 , ~~~~~~n=1,2,3,\ldots
\eeq
In Figure \ref{QuarkFit} we plot this fit.    
Turning to the massive  leptons,  i.e. the ``down" component of the SU(2) lepton doublets with masses 
$\{ m_e , m_\mu, m_\tau \}  \approx \{ 0.51, 105.0 , 1777.0 \} ~{\rm MeV } $,   
we find the fit 
\beq
\label{LeptonFitExp}
\mfamily_{n;{\rm leptons}} = \sqrt{\frac{\lambda_{\rm lep} }{3 \pi} }  M_{\rm lep} \, \,  e^{2 n  \lambda_{\rm lep}},  ~~~~ M_{\rm lep} = 0.92 \, {\rm MeV}, ~~~~ \lambda_{\rm lep}  = 1.4, ~~~~~ 
~n=1,2,3,\ldots
\eeq
See Figure \ref{LeptonFit}.   

\bigskip
A few comments are in order.      The above fits are for lower component of the SU(2) doublets of leptons verses the upper component for
the quarks,   and led to different RG periods $\lambda_{\rm lep},  \lambda_{\rm qu}$.        In the next section,  more reliable estimates of the RG periods $\lambda$ will continue to support this difference in $\lambda$ for upper verses lower components of these SU(2) doublets.    
For both cases  of leptons and  quarks the first family mass is underestimated.    This can be fixed with the angle $\vartheta_1$.  
However there may be other explanations based on the observation that the Russian Doll scaling is better at high energies,  which is already evident 
from \eqref{cyclicsG1} where only for large $n$ is $2 \cosh (n \lambda/2)  \approx e^{n \lambda/2}$.
Recent work on the Russian Doll BCS model indicates that the period of the RG can indeed be energy dependent \cite{Gorsky}.  
It is also interesting to note  that $ M_{\rm lep}$ and $ M_{\rm qu}$ are not so different in this crude fit.   

If there is a natural cut-off for the cyclic RG,   then only a finite number of families are realized up to this cut-off.  The natural cut-off here is clearly 
the electro-weak scale  of approximately $M_{\rm ew} = 2.5 \times 10^5$ MeV,   since the cyclic RG is only meaningful after SSB of the SU(2) to U(1).    If the cyclic RG only operates up to this electro-weak scale $M_{\rm ew}$,    then the number of families $n$ is the solution to
\beq
\label{NumFamQuark}
\sqrt{\frac{\lambda_{\rm qu}}{3 \pi} }\, M_{\rm qu} \,  e^{2 n \lambda_{\rm qu} } = M_{\rm ew}, ~~~~\Longrightarrow ~~ n = 3.14 ~~{\rm families}.
\eeq
Repeating the same calculation for $\lambda_{\rm lep}$ and $M_{\rm lep}$ gives $n=4.8$ potential families of leptons.  
Based on anomaly cancelation,  it's not possible to have an additional 4-th  family of leptons without also an additional family of quarks.   
Since  quark sector gives the strongest constraints  on a possible additional 4-th family in the Standard Model,   we interpret this as evidence for $N=3$ families.\footnote{
The  experimental constraints on a $4$-th family are primarily  inferred from the width of the $Z$-gauge boson,
the strongest being from the quark sector \cite{CMS}.      There are also strong theoretical constraints,    since the dominant contribution to the 
width of the $Z$ comes from the most massive top quark,  such that a more massive 4-th family could give larger contributions if just attributed to larger Yukawa coupling.     
    If we extrapolate the above fits to $n=4$,  then one expects a 4-th family top quark,  $t'$,   to have mass 
$ \mfamily_{t'} = \mfamily_{4;{\rm quarks}}  \approx 20 \, {\rm TeV} $.   This is  well above the experimental bound $m_{t'}  > 1.4 \, {\rm TeV}$. 
For a 4-th family lepton,   the above fit would place a 4-th family $\tau'$  at
$\mfamily_{\tau'} = \mfamily _{4;{\rm leptons}}  \approx 30 \, {\rm GeV} $.    The experimental bound here  is $m_{\tau'} > 100 \, {\rm GeV}$.   
A small increase of $\lambda_{\rm lep}$,  as little as $10\%$,   in our crude fit could easily place this over the experimental bound since the dependence on masses is exponential in $\lambda$.}

\subsection{Using  the  empirical Koide formula to constrain the RG period $\lambda$}

A more effective manner to determine the RG period $\lambda$ is to consider a quantities where the overall factor of $\sqrt{\frac{\lambda}{3 \pi} } M$ in \eqref{matterfit} cancels.     For $N$ families with masses $\mfamily_n, ~ n = 1, 2, 3, \ldots, N$,  let us define such a quantity:
\beq
\label{KN}
K ( \{ \mfamily \} ) = \frac{\sum_{n=1}^N \mfamily_n }{ \( \sum_{n=1}^N  \sqrt{\mfamily_n} \)^2 }
\eeq
Then based on \eqref{matterfit}, 
\beq
\label{KNlambda} 
K  (\lambda) =  \frac{ \sum_{n=0}^{N-1} \, e^{2 n \lambda} }{\( \sum_{n=0}^{N-1} \,  e^{n \lambda} \)^2}  =  \frac{(e^{2 N \lambda} -1 )(e^\lambda -1 )}{(e^{N \lambda} -1 )^2 
(e^\lambda +1 )}
\eeq
A non-cyclic RG actually corresponds to $\lambda = \infty$ since flows between the deep UV and IR typically take  an infinite RG ``time".   On the other hand,    if $\lambda$ is very small,   $K$ is a measure of $N$:
\beq
\label{KNlimits} 
\lim_{\lambda \to 0}  \, K (\lambda ) = \inv{N}, ~~~~~~ \lim_{\lambda \to \infty }  \, K (\lambda )  = 1.
\eeq

The above definition of $K$ was inspired by a curious formula first noticed by Koide \cite{Koide}.   
The masses of the bottom component leptons,  the  electron, muon, and tau,   are much more accurately known since they come from on-shell measured pole masses.    This is in contrast to the quark masses which receive radiative QCD corrections and these  masses are RG corrected to correspond to effective masses at an RG scale of about $2$\,GeV typically.         
Namely,  for the $N=3$   families of  the  heavy leptons,  electron, muon, and tau,  
\beq
\label{Kleptons} 
K_{\rm leptons}  =  \frac{m_e + m_\mu + m_\tau}{\( \sqrt{m_e}  + \sqrt{m_\mu}  +  \sqrt{m_\tau} \)^2}  = 0.666664~ \approx  \frac{2}{3} .
\eeq
For the masses indicated in Table \ref{KoideTable},   $K_{\rm leptons} = 0.666664$.     This observation remains a mysterious   phenomenological result with no compelling  theoretical explanation.

A cyclic RG can perhaps shed some light on this  Koide formula.          From \eqref{KNlambda},   solving for $\lambda$  based on \eqref{Kleptons} one finds $\lambda = 1.5668 \approx \pi/2$,  since $\pi/2 = 1.5708$.     This observation is as curious as the Koide formula itself.   
The question naturally arises:    Can this value of $\lambda \approx  \pi/2$  be explained by our cyclic RG flows?   
 Recall based on the above 1-loop RG,    
$\lambda = \pi/\sqrt{Q}$,   where $Q = g_1^2 - g_3^2$.  
Inspection of Figure \ref{Flows} shows that all the cyclic flows cross the $g_1$ axis with $g_3=0$.     Along this axis $Q= g_1^2$,   thus
$\lambda = \pi/2$ corresponds to the integer value  $g_1 = 2$.    Since this is based only on the 1-loop approximation and this $g_1$ is in a non-perturbative regime,  there is nothing particularly special about $g_1 = 2$ to 1-loop apart from being an integer.         Turning to higher orders,   
in the article \cite{RDolls2} the RG beta functions were explicitly computed up to 3 loops,   and this  led to a proposal for  non-perturbative extension of the RG invariant $Q$  
to all orders in the couplings $g_1, g_3$ for $\kappa =1$:
\beq
\label{Qallorders}
Q = \frac{g_1^2 - g_3^2}{(1 - g_3 /4)^2 (1-g_1^2 /16)} .
\eeq
First,  it was shown in \cite{RDolls2} that to higher orders  the RG period  is simply doubled in comparison with \eqref{period}:  $\lambda =  2 \pi/\sqrt{Q}$,  such that 
$\lambda = \pi/2$ corresponds to $Q=16$.      
Second,   the beta functions have an unexpected strong-weak duality,   namely under the transformation $g_{1,3} \to 16 /g_{1,3}$ the non-perturbative RG invariant $Q$ above  is invariant.    Thus $g_1 = 4$ is a special self dual point.   
For  $g_3 =0$, ~   $Q = g_1^2/(1-g_1^2/16)$,   such that $Q=16$ now corresponds to $g_1 = 2 \sqrt{2}$.   This is not yet a conclusive  argument for why $\lambda \approx  \pi/2$.   However it suggests that the values of $g_1, g_3$ that lead  to $\lambda \approx \pi/2$ are  closely tied to the self-dual points of these couplings,  and this is left as an open avenue for further investigations.  
    
Based on the above Koide formula it is interesting to consider the analogous $K$ for other fundamental particles,   the down quark components and the upper neutrino components of the SU(2) doublets.        
This data is presented in 
 Table \ref{KoideTable}.    One can see that the upper components of the SU(2) doublets have a higher $\lambda$ than the bottom components,   however  we cannot explain this based on our simplistic analysis.    These different $\lambda$ can perhaps be explained by the 
 angles $\vartheta_n$ in \eqref{vev4},  which for simplicity we have set all to $\pi/2$.

\begin{table}
\begin{center}
\begin{tabular}{||c||c||c|c||}
\hline\hline
${\rm Sector} $ & $ m_1,  ~~~ m_2, ~~~ m_3 ~~~ (\rm MeV) $  &   $K$    &   $\lambda$    \\
\hline\hline 
${\rm down ~leptons} $     &  $m_e =0.51099895, ~~ m_\mu = 105.6583755, ~~ m_\tau = 1776.93$   &  $0.666664 \approx 2/3$ &  $1.56679 \approx \pi/2$ \\
\hline
${\rm  down ~ quarks } $     &  $m_{\rm down}  = 4.7 , ~~~~ m_{\rm strange}  = 93.5 , ~~~~ m_{\rm bottom} = 4183. $   &  $0.7313$ &  $1.838$ \\
\hline
\hline
${\rm  up ~ quarks } $     &  $m_{\rm up}  = 2.16  , ~~~~ m_{\rm charm}  = 1273.  , ~~~~ m_{\rm top} =  172570. $   &  $0.84879$  &  $2.4969$ \\
\hline 
${\rm up~ leptons~ (neutrinos) } $     &  $m_{\nu_e}  < 4.5 \times 10^{-7}   , ~~~~ m_{\nu_\mu}  < 0.19   , ~~~~ m_{\nu_\tau}  <  18.2  $   &  $0.8315$  &  $2.377$ \\
\hline
\hline
\end{tabular}
\end{center}
\caption{Values of the Koide parameter $K$ and it's corresponding RG period $\lambda$ given in equations \eqref{KN} and \eqref{KNlambda} for various sectors with $N=3$ families.  ``Up"  verses ``down"  refers to upper verses lower components of the SU(2) doublets.   For the neutrinos,   we used the indicated upper limits to compute $\lambda$.}
\label{KoideTable}
\end{table}

\bigskip 
We close this section with additional  comments that are hard to resist,  and we ask the reader to allow us the liberty to elaborate.       Let  us re-emphasize that  we have not constructed a complete model that agrees with electro-weak  experiments and the above fits are extremely crude since they are based on only 3 data points.   

\bigskip

\medskip

\noindent (ii) ~  The fields $\Phi, \Phitilde$ comprise 8 real scalars.    As for the usual Higgs mechanism,   due to the breaking of SU(2) there are 3 
Goldstone bosons which lead to masses for the $W,Z$ gauge bosons.      This leaves 5 possible additional Higgs resonances.     
For a rough estimate,   the mass of an additional Higgs  could be around $m_{\rm{higgs'}} \sim M_{\rm Higgs} \,e^{2 \lambda}$.
Based on  the  above values for $\lambda_{\rm lep, qu}$,   if we take $\lambda \approx 2$,   then 
$m_{\rm{higgs'}} \approx 1 {\rm TeV}$,   which is also possibly within reach of colliders.   

\medskip

\medskip

\noindent (iii)  ~  In order to bolster our speculations,    we point out that it is well known that P.  Anderson anticipated the Higgs mechanism in explaining aspects of BCS superconductivity,  in particular how the Meissner effect can be attributed to the photon obtaining a mass \cite{Anderson}.     The Russian Doll BCS model \cite{LeClairSierraBCS} is a natural generalization of BCS theory with an additional coupling constant that breaks time-reversal symmetry and has a beta function that is identical to that in \eqref{NotFully1}.     

\medskip

\noindent (iv)  ~  We argued that our pseudo-hermitian model breaks $\CC\CP$ symmetry.      
   Since $\CC\CP$  breaking is thought to be necessary to explain the matter/anti-matter asymmetry in the Universe,    this could be a bonus unless the effect is too large.

\medskip
\noindent (v)  ~  The significance of the arbitrary angles $\vartheta_n$ in \eqref{vev4}  is not entirely clear.    In principle there is one such angle for each family.   They  are perhaps related to the mixing angles in the CKM matrix, but  this demands further study.   Incidently,   the CKM matrix for the Standard Model with 3 families  indeed has 3 independent mixing phases,  plus a $\CC\CP$ violating phase.

\section{Discussion and closing remarks}

The main result of this paper is the construction of  new kinds of  marginal operators for scalar fields in 4D  with SU(2) symmetry which have a rich structure of 
RG flows,  including cyclic flows,  when the SU(2) is broken to U(1),    and our focus has been on the latter.    Such operators make the hamiltonian pseudo-hermitian  $H^\dagger = \K H \K$,  and thus  formally non-unitary.   The non-unitarity is manifested in some negative norm states.   We showed that at low energies below the pair-production threshold,  or alternatively with a large
chemical potential,  2 body scattering is effectively unitary.     
These results are general and stand on their own,  and apply to other Lie groups.    They  could  have applications to the Standard Model of particle physics in the Higgs sector.     For instance,   two  SU(2) Higgs doublets can exhibit both SSB and a cyclic RG flow.     To support the validity of our interpretation,  we reviewed exact results for
QFT's in 2D with RG  beta functions that are identical those for our 2-Higgs models.     A general feature of cyclic theories is an infinite spectrum of 
physical masses or condensates with 
Russian Doll scaling.     We then took the liberty to speculate that a cyclic RG can perhaps resolve the hierarchy problem,
and furthermore,  could shed light on the  origin of families in the Standard Model,
   wherein  each additional family is identified with the next ``smaller"  and more massive  Russian Doll.    Based on a comparison with some 
   exactly solvable 2D models with identical RG beta functions and flows,   we argued that the more massive families should appear as resonances in the S-matrix.

\medskip

This work leads to many open questions,   and we list a few.

\medskip

$\bullet$~   
We defined the operator $\K$ by its  action on operators  in momentum space in \eqref{KU1},   but a physically more appealing interpretation is desirable.  
 The operator $\K$  acts non-locally on fields.       
The definition \eqref{KU1} requires $\vartheta = \pi$,  which is reminiscent of theories in 2D where topological charges at $\vartheta = \pi$ can strongly influence the critical behavior,    the classic example being  the O(3)  sigma model at $\vartheta = \pi$  \cite{Haldane}. 
In this regard,   the results in \cite{Wiegmann} might be relevant.\footnote{{\bf Note added:}  Based on our subsequent work to higher orders in perturbation theory  for our model and the massless RG flows we found between two different CFT fixed points \cite{RDolls2},    it was suggested that $\K$ can be interpreted as a topological defect,   analogous to a Verlinde line in 
2D.  In this 4D context,   one expects a codimension 2 defect,  i.e. a 2-dimensional surface.}

\medskip

$\bullet$~ All of our RG calculations were only carried out to only 1-loop.    For the 2D examples with the same 1-loop beta functions,
higher loop corrections confirmed the cyclic RG flow \cite{BLflow} for current-current perturbations to all orders.        Since the algebraic structure of our marginal operators closely parallel current-current perturbations in 2D,  we expect the cyclicity to hold to higher orders,  but this needs to be confirmed.\footnote{{\bf Note added:}  This was confirmed in \cite{RDolls2}.}

\medskip

$\bullet$~   There is some evidence for an underlying $\SL2Z$ structure in our model.    In \cite{RDolls2} we showed that the higher order beta functions have a $g \to 1/g$ duality,   like an S-duality $\tau \to -1/\tau$.      T-duality,  $\tau \to \tau + 1$ typically corresponds to a periodicity due to a $\vartheta$-angle,   as in $\CN=4$ supersymmetric Yang-Mills theory.      
 The cyclicity of the RG required us to identify 
$g_3 = \pm \infty$,  endowing the 2-parameter coupling constant space $(g_1, g_3)$ with the topology of a cylinder.
 This by itself does not establish an ${\rm SL}(2, \Z)$ symmetry,   since we have not precisely identified the elliptic modulus 
 $\tau$ as a function of the 2 couplings.    
The role of $\SL2Z$ is perhaps  clearer for the fully anisotropic case with the beta functions 
\eqref{FullyAniso} since integration of these 1-loop beta functions already leads to elliptic functions \cite{Elliptic}. 
Furthermore the 2D models with identical beta functions were proposed to have the S-matrix of Zamolodchikov 
\cite{Zelip}  built out of elliptic functions \cite{Elliptic}.    This S-matrix is characterized by a $\Z_4$ symmetry which is a subgroup of $\SL2Z$,  and furthermore,   it is known that this S-matrix transforms covariantly under $\SL2Z$.
In \cite{Elliptic},   the SU(2) broken to U(1) model,   the cyclic sine-Gordon model,   can be obtained from certain limits of Zamolodchikov's elliptic S-matrix,  and this is perhaps a way of understanding what is left of the $\SL2Z$ structure for our 
two parameter model. In recent years there have been interesting works that attempt to apply ${\rm SL}(2, \Z)$, or subgroups thereof,  to flavor physics.  Namely,  the Yukawa couplings are assumed to be modular forms for subgroups of
$\SL2Z$.    See \cite{Feruglio,ModularFlavor3,ModularFlavor2} and references therein.    These attempts are a priori not very well motivated, compared to  ${\rm SL}(2, \Z)$ symmetry in gauge theory couplings,  whereas flavor physics is intimately tied to Higgs physics.    
 The present work may provide such a motivation.

\medskip

$\bullet$~ We speculated that a cyclic RG flow with a signature of an infinite Russian Doll spectrum of vev's  $v_n$ may perhaps be at the origin of families in the 
Standard Model,  and made some very crude checks of whether this is plausible.   The strongest constraint on the fundamental 
RG period $\lambda$ came from the phenomenological Koide formula,  where just as curiously,    $\lambda \approx \pi/2$.    
If a cyclic RG is indeed valid below the electro-weak scale,   then based on our estimate of $\lambda$,   this allows for 3 RG cycles,  which would correspond to 3 families. 
  Clearly much more work is needed here to understand
 whether our model,  or variations of it,   can be made consistent with  all the stringent experimental constraints that are currently known.

\section{Acknowledgements} 

We wish to  thank  Yuval Grossman  for discussions on experimental bounds on a 4-th family.     We also with to  thank the late  Ken Wilson for early correspondence at the time of his last works \cite{GW1,GW2}.   
We also thank  Luis Alvarez-Gaum\'e, Denis Bernard,  Csaba Csaki,  Cheng-Yang Lee,    Ali  Mostafazadeh,  Michael Peskin,  and Germ\'an Sierra for discussions after the first version of this  work  appeared.

\end{document}